\documentclass[10pt, conference, compsocconf]{IEEEtran}
\IEEEoverridecommandlockouts
\usepackage{cite}

\ifCLASSINFOpdf
\else
\fi
\usepackage[nosumlimits,nonamelimits,cmex10]{amsmath}
\usepackage{array}

\usepackage[normalem]{ulem}			
\usepackage{proof}					
\usepackage{amssymb}				
\usepackage{stmaryrd,latexsym}		

\usepackage{amsthm}					
\theoremstyle{plain}
\newtheorem{thm}{Theorem}
\newtheorem{lem}[thm]{Lemma}
\newtheorem{prop}[thm]{Proposition}
\newtheorem{claim}[thm]{Claim}
\newtheorem{cor}[thm]{Corollary}

\theoremstyle{definition}
\newtheorem{defn}[thm]{Definition}

\newtheorem{expl}[thm]{Example}
\newtheorem{rem}[thm]{Remark}
\newtheorem{notn}[thm]{Notation}

\newtheorem{case}{Case}

\usepackage[matha]{mathabx}

\usepackage[final]{listings}		

\usepackage{optparams}				
\usepackage{xspace}

\usepackage[pdfpagemode={UseOutlines},bookmarks=true,bookmarksopen=true,
   bookmarksopenlevel=0,bookmarksnumbered=true,hypertexnames=false,
   colorlinks,linkcolor={blue},citecolor={blue},urlcolor={red},
   pdfstartview={FitV},unicode,breaklinks=true,final]{hyperref}

\usepackage[all]{xy}			

\usepackage{ifdraft}  				
\makeatletter
 \@drafttrue
 \@option@drafttrue
\makeatother

\usepackage[hide]{ed}

\makeatletter
\newcommand{\SG}[2][]{\ed@note{#2}{SG}{#1}}
\newcommand{\LS}[2][]{\ed@note{#2}{LS}{#1}}
\makeatother

\newcommand{\BC}{{\Cat C}}

\newcommand{\Set}{\Cat{Set}}

\newcommand{\Cat}{\mathbf}

\newcommand{\Opname}{\mathrm}

\newcommand{\Ob}{{\Opname{Ob}\,}}
\newcommand{\Hom}{\mathsf{Hom}}

\newcommand{\id}{\operatorname{id}}
\newcommand{\PSet}{{\mathcal P}}
\newcommand{\PSetN}{{\PSet^{\varstar}}}
\newcommand{\PSetNO}{{\PSet^{\varstar}_{\omega}}}
\newcommand{\PSetNOI}{{\PSet^{\varstar}_{\omega_1}}}

\newcommand{\impl}{\Rightarrow}

\newcommand{\consto}{\mathrel{\ooalign{$\circ$\cr\kern-1.65pt$\rightarrow$}}}

\newcommand{\nil}{\bot}

\newcommand{\comp}{}

\newcommand{\tensor}{\varotimes}

\newcommand{\mix}{\varstar} 
\newcommand{\tpair}[2]{{#1\wideparen{\;}\kern-1pt #2}}
\newcommand{\argument}{\_\!\_}

\newcommand{\downset}{\kern-2.8pt\downarrow\kern-2.8pt{}} 
\newcommand{\upset}{\kern-2.8pt\uparrow\kern-2.8pt{}} 

\newcommand{\unit}{\star}
\newcommand{\fst}{\operatorname{\sf fst}}
\newcommand{\snd}{\operatorname{\sf snd}}

\newcommand{\inl}{\operatorname{\sf inl}}
\newcommand{\inr}{\operatorname{\sf inr}}

\newcommand{\CASE}{\operatorname{\sf case}}
\newcommand{\OF}{\operatorname{\sf of}}
\renewcommand{\case}[3]{\CASE\kern1.2pt #1\kern1.2pt \OF\kern1.2pt #2;\kern1.2pt #3}

\newcommand{\caseOne}[2]{\CASE\kern1.2pt #1\kern1.2pt \OF\kern1.2pt #2}

\newcommand{\DO}{\operatorname{\sf do}}
\newcommand{\retOp}{\operatorname{\sf ret}}
\newcommand{\ret}[1]{\retOp #1}
\newcommand{\letTerm}[2]{\DO\kern1.2pt#1; #2}
\newcommand{\leteq}{\gets}

\newcommand{\letTermO}[2]{\DO_{\nu}\kern1.2pt#1; #2}
\newcommand{\LOOP}{\operatorname{\sf loop}}
\newcommand{\THEN}{\operatorname{\sf then}}

\makeatletter
\long\def\loopTerm@[#1][#2][#3]#4#5#6{
\DO\kern1.2pt#4 #1 \LOOP #3 #5 #2\THEN #6
}

\newcommand{\loopTerm}{
\optparams{\loopTerm@}{[;][\};][\{]}
}
\makeatother

\newcommand{\LET}{\operatorname{\sf init}}
\newcommand{\letLoop}[2]{\LET #1\kern1.2pt\LOOP\kern1.2pt\{#2\}}
\newcommand{\LetLoop}[2]{\LET #1\kern1.2pt\LOOP\kern1.2pt\bigl\{#2\bigr\}}

\newcommand{\IF}{\operatorname{\sf if}}
\newcommand{\ifTerm}[3]{\IF\kern1.2pt #1\kern4.2pt {\sf then}\kern2.2pt #2\kern4.2pt {\sf else}\kern2.2pt #3}
\newcommand{\ifTermO}[3]{\IF_{\nu} #1\kern2.2pt {\sf then}\kern2.2pt #2\kern2.2pt {\sf else}\kern2.2pt #3}

\newcommand{\INIT}{\operatorname{\sf init}}
\newcommand{\IN}{\operatorname{\sf in}}

\newcommand{\starTerm}[2]{\INIT #1 \IN #2^{\star}}

\newcommand{\WHILE}{\operatorname{\sf while}_{\nu}}
\newcommand{\whileTerm}[3]{\LET #1\kern1.2pt\WHILE\kern1.2pt #2 \kern2.2pt{\sf do}\kern2.2pt #3}
\newcommand{\whileTermS}[2]{\WHILE\kern1.2pt #1 \kern2.2pt{\sf do}\kern2.2pt #2}

\newcommand{\SEQ}{\operatorname{\sf seq}}
\newcommand{\seqTerm}[2]{\SEQ_{\nu}\kern1.2pt#1; #2}

\newcommand{\lrule}[3]{\textbf{(#1)}\;\;\infrule{#2}{#3}}

\newcommand{\infrule}[2]{\frac{#1}{#2}}

\newcommand{\brks}[1]{\langle #1\rangle}
\newcommand{\Brks}[1]{\bigl\langle #1\bigr\rangle}

\newcommand{\mypath}[1]{\llparenthesis\, #1\,\rrparenthesis}
\newcommand{\Path}[1]{\Bigl(\kern-2.1pt\Bigl|\, #1\,\Bigr|\kern-2.1pt\Bigr)}

\newlength{\myboxwidth}
{\end{center}\end{figure}}

\newcommand{\cn}{\mathsf{c}}
\newcommand{\uni}{\mathsf{u}}
\newcommand{\join}{\mathsf{U}}

\DeclareMathOperator{\redpar}{{
\declareslashed{}{
\vrule height2pt depth 2pt
\kern1pt
\vrule height2pt depth 2pt
}{0}{0}{\rightarrowtail}\slashed{\rightarrowtail}}}

\newcommand{\comma}{\mathbin{\mid}}

\newcommand{\klstar}{\star}  					

\renewcommand{\emptyset}{\varnothing} 

\newcommand{\Nat}{\mathbb{N}}

\newcommand{\poweq}{\approx}
\newcommand{\downup}[1]{\mathsf{cl}(#1)}
\newcommand{\op}{{\mathit{op}}}

\newenvironment{sparenumerate}%
{\begin{IEEEenumerate}[{\setlength{\leftmargin}{0pt}
    \setlength{\parsep}{0pt}%
    \setlength{\itemindent}{4ex}%
    \setlength{\itemsep}{2pt}\IEEEnocalcleftmargintrue}]}%
{\end{IEEEenumerate}}

{\begin{IEEEenumerate}[{\setlength{\leftmargin}{0pt}
    \setlength{\parsep}{0pt}%
    \setlength{\itemindent}{4ex}%
    \setlength{\itemsep}{2pt}\IEEEnocalcleftmargintrue}]}%
{\end{IEEEenumerate}}

\newenvironment{sparrmenumerate}%
{\begin{IEEEenumerate}[{\setlength{\leftmargin}{0pt}
    \setlength{\parsep}{0pt}%
    \setlength{\itemindent}{4ex}%
    \setlength{\itemsep}{2pt}\IEEEnocalcleftmargintrue}]}%
{\end{IEEEenumerate}}

\newenvironment{sparrmenumerateii}%
{\begin{IEEEenumerate}[{\setlength{\leftmargin}{10pt}
    \setlength{\parsep}{0pt}%
    \setlength{\itemindent}{4ex}%
    \setlength{\itemsep}{2pt}\IEEEnocalcleftmargintrue%
    \IEEEiedlabeljustifyl%
    \IEEEsetlabelwidth{iii)}}]}%
{\end{IEEEenumerate}}

\newenvironment{sparitemize}%
{\begin{IEEEitemize}[{\setlength{\leftmargin}{0pt}
    \setlength{\parsep}{0pt}%
    \setlength{\itemindent}{4ex}%
    \setlength{\itemsep}{2pt}\IEEEnocalcleftmargintrue}]}%
{\end{IEEEitemize}}

\renewcommand{\IEEEiedlistdecl}{\setlength{\IEEEiedtopsep}{-20pt}}

\newcommand{\myqed}{\hfill\IEEEQED}

        {\begin{enumerate}}%
        {\end{enumerate}}
        {\begin{enumerate}}%
        {\end{enumerate}}

\usepackage{ifpdf}					
\ifx\todo\undefined
\def\todo{}
\fi
\ifpdf
  \usepackage[usenames,dvipsnames]{color} 
  \renewcommand{\todo}[1]{
  {\color{red}{[TODO: #1]}}
  }
\else
  \usepackage[usenames,dvips]{color}
  \renewcommand{\todo}[1]{
  {\color{red}{[TODO: #1]}}
  }
\fi

\begin{document}
%
\title{Powermonads and Tensors of Unranked Effects}

\author{
\IEEEauthorblockN{Sergey Goncharov and Lutz Schr\"oder\thanks{Research supported by the DFG (project PLB, LU708/9-1).}}
\IEEEauthorblockA{\\Research Department Safe and Secure Cognitive Systems\\German Research Center for Artificial Intelligence (DFKI GmbH), Bremen, Germany\\
Email: Sergey.Goncharov@dfki.de, Lutz.Schroeder@dfki.de}
}


%


\maketitle

\begin{abstract}
  \noindent In semantics and in programming practice, algebraic
  concepts such as monads or, essentially equivalently, (large)
  Lawvere theories are a well-established tool for modelling generic
  side-effects. An important issue in this context are combination
  mechanisms for such algebraic effects, which allow for the modular
  design of programming languages and verification logics. The most
  basic combination operators are sum and tensor: while the sum of
  effects is just their non-interacting union, the tensor imposes
  commutation of effects. However, for effects with unbounded arities,
  these combinations need not in general exist. Here, we introduce the
  class of uniform effects, which includes unbounded nondeterminism
  and continuations, and prove that the tensor does always exist if
  one of the component effects is uniform, thus in particular
  improving on previous results on tensoring with continuations. We
  then treat the case of nondeterminism in more detail, and give an
  order-theoretic characterization of effects for which tensoring with
  nondeterminism is conservative, thus enabling nondeterministic
  arguments such as a generic version of the Fischer-Ladner encoding
  of control operators.
\end{abstract}


%

\renewcommand{\IEEEQED}{\IEEEQEDopen}

\section{Introduction}
\noindent Both in actual programming languages and in their semantics
and meta-theory, one encounters a wide variety of phenomena that can
be subsumed under a broadly understood notion of side-effect, such as
various forms of state, input/output, resumptions, backtracking,
nondeterminism, continuations, and many more. This proliferation of
effects motivates the search for generic frameworks that encapsulate
the exact nature of side-effects and support abstract formulations of
programs (such as Haskell's generic while-loop), semantic principles,
and program logics. A fairly well-established abstraction of this kind
is the modelling of side-effects as monads, following seminal work by
Moggi~\cite{Moggi91}; this principle is widely used in programming
language semantics
(e.g.~\cite{JacobsPoll03,Papaspyrou01,ShinwellPitts05,Harrison06}) and
moreover underlies the incorporation of side-effects in the functional
programming language Haskell~\cite{Wadler97}. Besides supporting
generic results that can be instantiated to particular effects at
little or no cost, monads allow for a clear delineation of the scope
of effects~\cite{MoggiSabry01}.  A more recent development is the
advancement of Lawvere theories~\cite{Lawvere63} for the generic
modelling of effects, thus emphasizing their algebraic
nature~\cite{PlotkinPower02}.

One advantage of these approaches is that they provide for a
\emph{modular} semantics of effects. It has been observed that many
effects, such as state, exceptions, and continuations, induce
so-called \emph{monad transformers} that can be seen as adding the
respective effect to a given set of
effects~\cite{CenciarelliMoggi93,LiangHudakAtAl95}; again, the notion
of monad transformer plays a central role in Haskell. More recently,
it has been shown that many monad transformers arise from binary
combination operators that join effects in a prescribed way. The most
important among these constructions are the \emph{sum} of effects,
which corresponds simply to the disjoint union of algebraic theories,
and the \emph{tensor}, which additionally imposes a commutation
condition~\cite{HylandLevyAtAl07,HylandPlotkinAtAl06}. E.g., the
exception monad transformer is summation with the exception monad, and
the state monad transformer is tensoring with the state
monad~\cite{PowerShkaravska04}. These combination methods are often
mixed; e.g.~\cite{Stark08} uses both sums and tensors of
nondeterminism with other effects. (Previous work on the specific
combination of unbounded nondeterminism and probabilistic choice uses
a different form of interaction than imposed by the
tensor~\cite{TixKeimelAtAl09,VaraccaWinskel06}.)

One of the problems that arise with sum and tensor in the context of
\emph{large} Lawvere theories, i.e.\ theories that can be
\emph{unranked} in the sense that their operations have unbounded
arities, such as unbounded nondeterminism or continuations, is that
for reasons of size, the combined theories need not exist in general;
e.g.\ we show in recent work~\cite{GoncharovSchroder11} that tensors
of unranked theories with the theory of lists may fail to exist. In
the present work, we introduce the notion of \emph{uniform} theory,
and prove that the tensor of two large Lawvere theories always exists
if one of them is uniform. The class of uniform theories includes
several variants of nondeterminism (e.g.\ unbounded and countable, but
not finite) as well as, somewhat surprisingly, continuations; thus,
our existence result improves a previous result stating that the
tensor of any \emph{ranked} theory with continuations always
exists~\cite{HylandLevyAtAl07}.

One may read this result as yielding a number of new monad
transformers; we are particularly interested in nondeterminism monad
transformers, which we dub \emph{powermonads}. This leads us to a
second problem associated specifically with the tensor: since the
tensor imposes a complex algebraic interaction between the component
effects, it cannot in general be expected to be \emph{conservative} in
the sense that the components embed into the tensor.

To deal with this issue in the special case of nondeterminism, we
focus on \emph{bounded} theories $L$, which come with a natural
approximation ordering. We begin by giving a simplified construction
of tensoring with nondeterminism, which is informed by but technically
independent of the general existence result (and, e.g., applies also
to tensoring finitary theories with finite nondeterminism, although
the latter fails to be uniform): morphisms in the tensor of $L$ with
nondeterminism are sets of $L$-morphisms modulo \emph{rectangular
  equivalence}, a comparatively simple equivalence that forces
uniqueness of tupling morphisms. From there, we obtain a more
order-theoretic description of the tensor in terms of \emph{closed}
sets of $L$-morphisms, which leads to a simple characterization of
theories for which tensoring with nondeterminism is
order-theoretically conservative.


The main reason for our interest in tensoring with nondeterminism is
that it yields exactly the free extension of a given theory to a
\emph{completely additive} theory, i.e.\ one that is enriched over
complete join semi-lattices; this amounts to having choice operators
that distribute over sequential composition on both sides (hence
providing a trace-based rather than a bisimulation-based
perspective). Thus, whenever a theory $L$ can be conservatively
tensored with nondeterminism, one can conduct equational and
order-theoretic proofs in it pretending that $L$ is completely
additive. E.g., one can use the well-known translation of imperative
constructs~\cite{FischerLadner79}
\begin{align*}
\mathtt{if}~b~\mathtt{then}~p~\mathtt{else}~q 	&:= b?; p + (\neg b)?; q\\
\mathtt{while}~b~\mathtt{do}~p 					&:= (b?;p)^{\star};(\neg b)?
\end{align*}
that we dub the Fischer-Ladner encoding \emph{generically}, i.e.\ for
any effect satisfying our conservativity conditions. Besides
simplifying the reasoning, this uncovers the nondeterministic flavour
of imperative branching~\cite{Plotkin77}.

The material is organized as follows. We recall basic facts on monads
and Lawvere theories in Section~\ref{sec:lawvere}. In
Section~\ref{sec:tensors}, we review tensor products, and proceed
directly to the main existence result for tensors with uniform
theories. We discuss additive theories in Section~\ref{sec:additive},
and present our results on conservativity of tensoring with
nondeterminism in Section~\ref{sec:nondet-tensor}.

\section{Large Lawvere Theories and Monads}\label{sec:lawvere}

\noindent In a nutshell, the principle of monadic encapsulation of
side-effects originally due to Moggi~\cite{Moggi91} and subsequently
introduced into the functional programming language Haskell as the
principal means of dealing with impure features~\cite{Wadler97}
consists in moving the side effect from the function arrow into the
result type of a function: a side-effecting function $X\to Y$ becomes
a pure function $X\to TY$, where $TY$ is a type of side-effecting
computations over $Y$; the base example is $TY=S\to (S\times Y)$ for a
fixed set $S$ of states, so that functions $X\to TY$ are functions
that may read and update a global state (more examples will be given
later). Formally, a \emph{monad} on the category of sets, presented as
a \emph{Kleisli triple} $\mathbb{T}=(T,\eta,\argument^{\klstar})$, consists of
a function $T$ mapping sets $X$ (of values) to sets $TX$ (of
computations), a family of functions $\eta_X:X\to TX$, and a map
assigning to every function $f:X\to TY$ a function $f^{\klstar}:TX\to TY$ that
lifts $f$ from $X$ to computations over $X$. These data are subject to
the equations
\begin{math}
  \eta^{\klstar}=\id, f^{\klstar}\eta=f, (f^{\klstar} g)^{\klstar}=f^{\klstar}g^{\klstar},
\end{math}
which ensure that the \emph{Kleisli category} of $\mathbb{T}$, which
has sets as objects and maps $X\to TY$ as morphisms, is actually a
category, with identities $\eta:X\to TY$ and composition $f^{\klstar}g$. On
$\Set$, all monads are \emph{strong}, i.e.\ equipped with a natural
transformation $X\times TY\to T(X\times Y)$ satisfying a number of
coherence conditions~\cite{Moggi91}.

Monads were originally intended as abstract presentations of algebraic
theories, with $TX$ abstracting the free algebra over $X$, i.e.\ terms
over $X$ modulo provable equality. It has been shown that the
algebraic view of monads gives rise to computationally natural
operations for effects; e.g.\ the state monad (with state set $S=V^L$
for sets $V$ of values and $L$ of locations) can be algebraically
presented in terms of operations $\mathit{lookup}$ and
$\mathit{update}$~\cite{PlotkinPower02}. Categorically, this shift of
viewpoint amounts to generating monads from Lawvere theories. To cover
unranked theories, we use the notion of large Lawvere
theory~\cite{Dubuc70}, introduced into the theory of generic effects
in~\cite{HylandLevyAtAl07}. Generally, we denote hom-sets of a
category $\BC$ in the form $\BC(A,B)$.
\begin{defn}[Large Lawvere theory]\label{def:lawvere}
  A \emph{large Lawvere theory} is given by a locally small category
  $L$ with small products, together with a strict product preserving
  identity-on-objects functor $I: \Set^\op\to L$. 
  We call $I$ the \emph{indexing
    functor}, and we denote $If$ by $[f]$ for a map $f$. A morphism of
  large
  Lawvere theories $L_1\to L_2$ is a functor $L_1\to L_2$ that
  commutes with the indexing functors (and hence preserves small
  products). A \emph{model} of a large Lawvere theory $L$
  in a category $\BC$ with small products is a small product
  preserving functor $L\to\BC$.
\end{defn}
\noindent The algebraic intuition behind these definitions is that the
objects of a large Lawvere theory are sets $n,m,k,\dots$ of variables,
and morphisms $n\to m$ are $m$-tuples of terms over $n$, or
substitutions from $m$ into terms over $n$. The indexing functor
prescribes the effect of rearranging variables in terms.  The notion
of model recalled above implies that Lawvere theories provide a
representation of effects that is independent of the base category
$\BC$, and given enough structure on $\BC$, a Lawvere theory will
induce a monad on $\BC$.  E.g., in categories of domains, the theory
of finite non-blocking nondeterminism
(Example~\ref{expl:theories}.\ref{item:non-det} below) induces
precisely the Plotkin powerdomain monad (while the Hoare and Smyth
powerdomains require enriched Lawvere theories)~\cite{AbramskyJung94}.

It is well-known that large Lawvere theories and strong monads on
$\Set$ form equivalent (overlarge)
categories~\cite{Dubuc70,HylandLevyAtAl07}. The equivalence maps a
large Lawvere theory $L$ to the monad $T_LX=L(X,1)$ (we elide the full
description), and a monad $T$ to the dual of its Kleisli category. We
therefore largely drop the distinction between monads and large
Lawvere theories, and freely transfer concepts and examples from one
setting to the other; occasionally we leave the choice open by just
using the term \emph{effect}.

We say that a large Lawvere theory $L$ is \emph{ranked} if it can be
presented by operations (and equations) of arity less than $\kappa$
for some cardinal $\kappa$; otherwise, $L$ is
\emph{unranked}. Categorically, $L$ having rank $\kappa$ amounts to
preservation of $\kappa$-directed colimits by the induced monad. If
$L$ has rank $\kappa$, then $L$ is determined by its full subcategory
spanned by the sets of cardinality less than $\kappa$. If $L$ has rank
$\omega$, we say that $L$ is \emph{finitary}.


\begin{expl}\label{expl:theories}
  \begin{sparenumerate}
  \item \emph{Global state:} as stated initially, $TX=S\to(S\times X)$
    is a monad (for this and other standard examples, we omit the
    description of the remaining data), the well-known \emph{state
      monad}. A variant is the \emph{partial state monad}
    $TX=S\to(S\times X)_\bot$, where $X_\bot$ extends $X$ by a fresh
    element $\bot$ representing non-termination. (This induces a
    relational model of non-termination in the spirit of PDL and
    related formalisms; a domain-theoretic treatment of
    non-termination requires a domain-enriched Lawvere theory in which
    $\bot$ is explicitly a bottom element).
  \item\label{item:non-det} \emph{Nondeterminism:} the \emph{unranked}
    large Lawvere theory $L_\PSet$ for nondeterminism arises from the
    powerset monad $\PSet$. It has $m$-tuples of subsets of $n$ as
    morphisms $n\to m$. Variants arise on the one hand by restricting
    to nonempty subsets, thus ruling out non-termination, and on the
    other hand by bounding the cardinality of subsets. We denote
    nonemptyness by a superscript $\varstar$, and cardinality bounds
    by subscripts. E.g., the large Lawvere theory $L_{\PSetNO}$
    describes finite non-blocking nondeterminism; its morphisms $n\to
    m$ are $m$-tuples of nonempty finite subsets of $n$.
    Yet another variant arises by replacing sets with multisets, i.e.\
    maps $X\to(\Nat\cup\{\infty\})$, thus modelling weighted
    nondeterminism~\cite{DrosteKuichAtAl09} as a large Lawvere theory
    $L_{\mathit{mult}}$.
  \item\label{item:cont} \emph{Continuations:} The continuation monad
    maps a set $X$ to the set $(X\to R)\to R$, for a fixed set $R$ of
    results. The corresponding \emph{unranked} large Lawvere theory
    $L^R_{\mathit{cont}}$ has maps $m\to((n\to R)\to R)$ as morphisms
    $n\to m$.
  \item \emph{Input/Output:} For a given set $I$ of input symbols, the
    Lawvere theory $L_I$ for input is generated by a single $I$-ary
    operation $\mathit{in}$; it is an \emph{absolutely free} theory,
    i.e.\ has no equations. Similarly, given a set $O$ of output
    symbols, the Lawvere theory $L_O$ for output is generated by unary
    operations $\mathit{out}_o$ for $o\in O$.
  \end{sparenumerate}
  \noindent Further effects that fit the algebraic framework are
  exceptions ($TX=X+E$), resumptions ($RX=\mu Y.\,T(X+Y)$ for a given
  base effect $T$) and many more.
\end{expl}

\begin{notn}
  Let $L$ be a large Lawvere theory.
  For an object $n$ of $L$ and $i\in n$, we let $\kappa_i$ denote the
  map $1\to n$ that picks $i$. Thus, the $\kappa_i$ induce product
  projections $[\kappa_i]:n\to 1$ in $L$. Given two sets $n$ and $m$,
  their $\Set$-product $n\times m$ can be viewed as the sum of $m$
  copies of $n$ in $\Set$, and hence as the $m$-th power of $n$ in
  $L$. This induces for every $f:n\to k$ in $L$ the morphisms
  $f\tensor m:n\times m\to k\times m$ and $m\tensor f:m\times n\to
  m\times k$.
\end{notn}
\noindent A convenient way of denoting generic computations is the
so-called \emph{computational metalanguage}~\cite{Moggi91},
which has found its way into functional programming in the shape of
Haskell's do-notation. We briefly outline the version of the
metalanguage we use below.

The metalanguage serves to denote morphisms in the underlying category
of a given monad, using the monadic structure; since large Lawvere
theories correspond to monads on $\Set$, the metalanguage just denotes
maps in our setting. We let a signature $\Sigma$ consist of a set
$\mathcal{B}$ of \emph{base types}, to be interpreted as sets, and a
collection of typed \emph{function symbols} to be interpreted as
functions. Here, we assume that \emph{types} $A,B\in\mathcal{T}$ are
generated from the base types by the grammar
\begin{equation*}
  A,B::=1\mid A\in\mathcal{B}\mid A+B\mid A\times B\mid TA
\end{equation*}
where $+$ and $\times$ are interpreted as set theoretic sum and
product, respectively, $1$ is a singleton set, and $T$ is application
of the given monad. We then have standard formation rules for
terms-in-context $\Gamma\rhd t:A$, read `term $t$ has type $A$ in
context $\Gamma$', where a \emph{context} is a list
$\Gamma=(x_1:A_1,\dots,x_n:A_n)$ of typed variables (later, contexts
will mostly be omitted):
\begin{gather*}
\infrule{x:A\in\Gamma}{\Gamma \rhd x:A}\quad
\infrule{f:A\to B\in\Sigma ~~~ \Gamma\rhd t:A}
{\Gamma\rhd f(t):B}\quad
\infrule{}
{\Gamma\rhd\unit:1}\\[1ex]
\infrule{\Gamma\rhd t:A ~~~~
\Gamma\rhd u:B}
{\Gamma\rhd \langle t,u\rangle : A \times B}\quad
\infrule{\Gamma\rhd t:A\times B}
{\Gamma\rhd\fst t:B~~~\Gamma\rhd\snd t:B}\\[1ex]
\infrule{\Gamma\rhd s:A+B ~~~~ \begin{array}[b]{l}\Gamma,x:A\rhd t:C\\\Gamma,y:B\rhd u:C\end{array}}
{\Gamma\rhd\case{s}{\inl x\mapsto t}{\inr y\mapsto u}:C}\\[1ex]
\infrule{\Gamma\rhd t:A}
{\Gamma\rhd\inl t:A+B}\qquad
\infrule{\Gamma\rhd t:B}
{\Gamma\rhd\inr t:A+B}
\end{gather*}
This syntax supports, e.g., the standard encoding of the if-operator
as
\begin{equation*}\label{eq:if_def}
  \ifTerm{b}{p}{q}=\case{b}{\inl\unit\mapsto p}{\inr\unit\mapsto q},
\end{equation*}
for $b:2$, where $2=1+1$.
Beyond this, we have monadic term constructors
\begin{equation*}
\infrule{\Gamma \rhd t:A}
{\Gamma \rhd \ret{t}:TA}\quad
\infrule{\Gamma\rhd p:TA ~~
\Gamma, x:A\rhd q:TB}
{\Gamma\rhd \letTerm{x\leteq p}{q}:TB}
\end{equation*}
called~\emph{return} and~\emph{binding}, respectively. Return is
interpreted by the unit $\eta$ of the monad, and can be thought of as
returning a value. A binding $\letTerm{x\leteq p}{q}$ executes $p$,
binds its result to $x$, and then executes $q$, which may use $x$ (if
not, mention of $x$ may be omitted). It is interpreted using Kleisli
composition and strength, where the latter serves to propagate the
context $\Gamma$~\cite{Moggi91}. In consequence, one has the
\emph{monad laws}
\begin{gather*}
\letTerm{x\leteq p}{\ret{x}} = p\qquad
\letTerm{x\leteq\ret{a}}{p} = p[a/x]\\
\letTerm{x\leteq (\letTerm{y\leteq p}{q})}{r} = \letTerm{x\leteq p;y\leteq q}{r}
\end{gather*}
Terms of a type $T A$ are called~\emph{programs}. 

\section{Tensors of Large Lawvere Theories}
\label{sec:tensors}

\noindent One of the key benefits of the monadic modelling of effects
is that it allows for a modular treatment, where effects are combined
from basic building blocks according to the demands of the programming
task at hand. In current programming practice (specifically in
Haskell~\cite{Peyton-Jones03}), this is typically achieved by
generalizing a given effect to a \emph{monad
  transformer}~\cite{Moggi91,LiangHudakAtAl95}, i.e.\ a function that
maps monads to monads, in the process extending them with a given
effect. For instance, the \emph{state monad transformer} $\mathit{ST}$
for a given set $S$ of states maps a given monad $T$ to the monad
$\mathit{ST}(T)$ with $\mathit{ST}(T)(X)=S\to T(S\times X)$. Monad
transformers are very general, but do not support a great deal of
meta-theoretic results, as no further properties are imposed on them;
e.g., they need not be functorial. It has been shown
in~\cite{HylandPlotkinAtAl06} that many monad transformers arise from
a few basic binary operations on Lawvere theories (equivalently on
monads). E.g., the exception monad transformer, which maps a monad $T$
to the monad $T(\argument+E)$ for a fixed set $E$ of exceptions, is
just summation with $\argument+E$; expressed in terms of large Lawvere
theories, the sum $L_1+L_2$ of two effects $L_1$, $L_2$ is simply the
disjoint union of the associated theories, i.e.\ is universal w.r.t.\
having morphisms $L_1\to L_1+L_2\leftarrow L_2$. Another important
operation is the \emph{tensor} which additionally imposes a strong
form of interaction between the component theories in the form of a
commutation law.
\begin{defn}[Tensor]\cite{HylandLevyAtAl07}\label{def:lawvere_tensor}
  The \emph{tensor} ${L_1\tensor L_2}$ of large Lawvere theories
  $L_1,L_2$ is the large Lawvere theory which is universal w.r.t.\
  having \emph{commuting} morphisms $L_1\to L_1\tensor L_2\leftarrow
  L_2$ (elided in the notation), if such a universal theory exists.
  Here, commutation is satisfaction of the \emph{tensor law}, i.e.\
  given $f_1: n_1\to m_1$ in $L_1$ and $f_2: n_2\to m_2$ in $L_2$ we
  demand commutativity of the diagram
\begin{equation*}
    \xymatrix@R30pt@C30pt@M6pt{
      n_1\times n_2	\ar[r]^{n_1\tensor f_2} \ar[d]_{f_1\tensor n_2} & n_1\times m_2\ar[d]^{f_1\tensor m_2}\\
      m_1\times n_2	\ar[r]^{m_1\tensor f_2} &m_1\times m_2.
    }
\end{equation*}
\end{defn}
\noindent By the equivalence between large Lawvere theories and
monads, this induces also a notion of \emph{tensor of
  monads}~\cite{HylandLevyAtAl07}. The computational meaning of the
commutation condition becomes clearer in the computational
metalanguage: if we extended the metalanguage with subtypes $T_iA$ of
$TA$ interpreted using the component monads $T_1$, $T_2$ of the tensor
$T=T_1\tensor T_2$, it amounts to the equality
\begin{multline*}
  \letTerm{x_1\leteq p_1;x_2\leteq p_2}{\ret\brks{x_1,x_2}}=\\
  \letTerm{x_2\leteq p_2;x_1\leteq p_1}{\ret\brks{x_1,x_2}}
\end{multline*}in context $\Gamma_1,\Gamma_2$, where $\Gamma_i\rhd
p_i:T_iA_i$ for $i=1,2$; i.e.\ programs having only effects from $T_1$
do not interfere with programs having only effects from $T_2$.
\begin{expl}\cite{PowerShkaravska04}
  Tensoring with the state monad $TX=S\to (S\times X)$ yields exactly
  the standard state monad transformer (in particular, tensors with
  $T$ always exist).
\end{expl}
\noindent Sum and tensor of large Lawvere theories need not exist in
general.  This is a size issue --- if arities of operations are
unbounded, then the terms over a given set of variables need not form
a set. E.g., the sum $L_1+L_2$ of almost any unranked large Lawvere
theory $L_1$ and the theory $L_2$ generated by a single unary
operation and no equations fails to
exist~\cite{HylandPower07}. Generally, the tensor has a better chance
to exist than the sum, since it introduces additional equations, and
in fact existence of the sum implies existence of the
tensor~\cite{HylandLevyAtAl07}.  Nevertheless, the tensor of two large
Lawvere theories may fail to exist even when one of the component
theories is ranked (of course, it does exist in case \emph{both}
components are ranked); e.g.\ there are theories whose tensor with the
list theory fails to exist~\cite{GoncharovSchroder11}. We proceed to
show that the tensor exists whenever one of the component theories is
uniform in the sense defined presently.

\begin{defn}[Uniformity]\label{def:lawvere_contr}
  Let $L$ be a large Lawvere theory. The \emph{constants} of $L$ are
  the elements of $c_L:=L(0,1)$. For every set $n$ we denote by
  $\cn^n_L:n\to n+c_L$ the morphism $[\id]\times\prod_{f\in c_L} f$.
  We say that $L$ is \emph{uniform} if for every $L$-morphism $f:n\to
  m$ there exists a \emph{generic morphism}, i.e.\ a morphism $\hat
  f:k\to 1$ for some set $k$ such that there exists a set-function
  $\uni:k\times m\to n+c_L$ with $f=(\hat f\tensor m)\comp
  [\uni]\comp\cn^n_L$.
\end{defn}
\noindent In other words, a theory is uniform if all terms over a
given set $n$ of variables can be obtained from a single generic term
$\hat f$, possibly having more variables, by substituting for the
variables of $\hat f$ either variables from $n$ or constants. The
relevance to existence of tensors is clear: if a theory $L_2$ is
uniform, then the tensor law of a putative tensor $L_1\tensor L_2$ can
always be made to apply to a term that has, say, a top layer of
operations from $L_1$ whose arguments have a top layer from $L_2$.
\begin{rem}\label{rem:k-bound}
  It is easy to see that in Definition~\ref{def:lawvere_contr}, $k$
  can be bounded by $(n+c_L)^m$.
\end{rem}

\begin{expl}
  \begin{sparenumerate}
  \item \emph{The theory $L_\PSetN$ of non-blocking unbounded
      nondeterminism is uniform:} Recall that a morphism $f:n\to m$
    in $L_\PSetN$ is a family of $m$ nonempty subsets of $n$. As a
    generic morphism $\hat f$ for $f$, we can thus take the full set
    $n$, seen as a morphism $n\to 1$, from which any other subset of
    $n$ can be obtained by identifying some of the variables.
  \item \emph{The theory $L_\PSet$ of unbounded nondeterminism is
      uniform:} The argument is analogous as for $L_\PSetN$, except
    that we now need to use also the constant $\emptyset$ in
    substitutions in order to obtain the empty set as a substitution
    instance of the generic morphism $\hat f$.
  \item \emph{The theory $L_{\PSet_\omega}$ of finite nondeterminism
      fails to be uniform:} if $\sup_{i\in m}|A_i|=\infty$ for an
    infinite family $(A_i)_{i\in m}$ of finite subsets of an infinite
    set $n$, then there is no single finite set from which all sets
    $A_i$ can be obtained by substituting $\emptyset$ or variables
    from $n$.
  \item \emph{The theory $L_{\PSet_{{\omega_1}}}$ of countable
      nondeterminism is uniform:} any infinite countable subset of
    $n$ will serve as a generic morphism $\hat f$ for any morphism
    $f:n\to m$, i.e.\ any family of at $m$ most countable subsets of
    $n$.
  \item \emph{The theory $L_{\mathit{mult}}$ of unbounded weighted
      nondeterminism is uniform}: Recall from
    Example~\ref{expl:theories}.\ref{item:non-det} that a morphism
    $f:n\to m$ in $L_{\mathit{mult}}$ is a family of $m$ multisets
    over $n$ (i.e.\ maps $n\to\Nat\cup\{\infty\}$). As a generic
    morphism $\hat f$, we can take the multiset over $\Nat\times n$
    that contains every element with multiplicity $1$.
  \end{sparenumerate}
\end{expl}
\noindent Moreover, uniformity also subsumes continuations, a fact
that we state and prove separately:
%
\begin{lem}
  For every $R$, the continuation theory $L^R_{\mathit{cont}}$
  (Example~\ref{expl:theories}.\ref{item:cont}) is uniform.
\end{lem}
\begin{IEEEproofNoQED}
  W.l.o.g.\ $|R|\geq 2$. We identify the set of constants of
  $L^R_{\mathit{cont}}$ with $R$. Let $f:n\to m$ in
  $L^R_{\mathit{cont}}$; recall that $L^R_{\mathit{cont}}$ is the dual
  of the Kleisli category of the continuations monad, i.e.\ $f$ is a
  map $m\to ((n\to R)\to R)$.
  Pick $J$ such that $|m|\le |R^J|$; we can assume w.l.o.g.\ that
  $m=R^J$, as we can just pad out $f:n\to m$, thought of as a family
  of $m$ morphisms $n\to 1$, with sufficiently many copies of one of
  these morphisms.
  The required generic morphism for $f$ is $\hat f:n+J\to 1$, defined by
\begin{displaymath}
  \hat f(c) = f(\lambda j.\,c(\inr j))(\lambda a.\,c(\inl a)) 
\end{displaymath} 
for $c:n+J\to R$: Let $\uni:(n+J)\times m\to n+R$, 
 $$\uni(x,i)=\case{x}{\inl y\mapsto\inl y}{\inr
  j\mapsto \inr i(j)}.$$ Then for $i\in m=R^J$ and $k:n\to R$,
\begin{align*}
   \bigl((\hat f&\tensor m)[\uni]\cn^n_L\bigr)(i)(k) \\
  & = \hat f(\lambda x.\,\case{u(x,i)}{\inl y\mapsto k(y)}{\inr r\mapsto r})\\
  & = \hat f(\lambda x.\,\case{x}{\inl y\mapsto k(y)}{\inr j\mapsto i(j)})\\
  & = f(i)(k).\\[-3.2em]
\end{align*}
\myqed
\end{IEEEproofNoQED}%

\vspace{1ex}
\noindent The main existence result for tensors is as follows.
\begin{thm}\label{thm:contr}
  Let $L_1,L_2$ be large Lawvere theories, and let $L_2$ be
  uniform. Then the tensor product $L_1\tensor L_2$ exists.
\end{thm}
\begin{IEEEproof}[Proof sketch]
  By explicit syntactic construction of the tensor product
  ${L=L_1\tensor L_2}$.  One constructs a precursor $\BC$ of the
  tensor whose morphisms $n\to m$ are equivalence classes of paths
  $n\to m$; a single step $k\to l$ in a path is of the form $f\mix g$
  where $f:p\to l$ in $L_2$ and $g:k\to p$ in $L_1$. The equivalence
  is the congruence $\sim$ on paths $\mypath{f_1\mix
    g_1\comma\dots\comma f_u\mix g_u}$ generated by $\mypath{[\id]\mix
    [\id]}\sim\mypath{}$, $\mypath{f[e]\mix g}\sim\mypath{f\mix [e] g}$, and
\begin{multline*}
  \mypath{f\mix (n'\tensor g)\comma (f'\tensor m)\mix g'}\sim\mypath{f (f'\tensor m')\mix (n\tensor g) g'}.
\end{multline*}
Using uniformity of $L_2$, one shows that every morphism of $\BC$ has
a representative of the form
\begin{equation}\label{eq:seq_norm_body}
\mypath{f\mix g\comma\cn^n_{L_2}\mix[\id]}
\end{equation}
(recall notation from Definition~\ref{def:lawvere_contr}).  One shows
moreover that in \eqref{eq:seq_norm_body}, the domain of $f$ can be
taken to be $k=L_1(n+c_{L_2},1)$, so that $\BC$ is locally small. One
defines a functor $I:\Set^\op\to\BC$ by $Ie=[e]\mix[\id]$. It turns
out that $I$ maps products to \emph{weak} products, i.e.\
factorizations through the product exist but need not be unique; this
is amended by further quotienting.
\end{IEEEproof}
\begin{cor}
  For a large Lawvere theory $L_1$, the tensor $L_1\tensor L_2$ exists
  if $L_2$ is one of the following theories:
  \begin{IEEEitemize}
  \item   unbounded or countable nondeterminism $L_\PSet$,
    $L_{\PSet_{{\omega_1}}}$;
  \item unbounded or countable non-blocking nondeterminism $L_{\PSetN}$, $L_{\PSetNOI}$;
  \item weighted nondeterminism $L_{\mathit{mult}}$; or
  \item continuations $L^R_{\mathit{cont}}$.
  \end{IEEEitemize}
\end{cor}
\noindent Of course, a corresponding result holds for monads. This
result induces new monad transformers for nondeterminism,
continuations, etc. The existence result for tensoring with
continuations improves over previous results stating that the tensor
of continuations with any \emph{ranked} theory
exists~\cite{HylandLevyAtAl07}. The results involving nondeterminism
are, to our knowledge, entirely new. We refer to tensoring with
any form of nondeterminism as a \emph{powermonad} construction.

%
%
%
%
%
%
%
%
%
%
\section{Completely Additive Monads\\and the Fischer-Ladner Encoding}\label{sec:additive}

\noindent Having shown that the tensor of any effect with
nondeterminism always exists, we proceed to show that this amounts to
a universal construction of an \emph{additive} theory, i.e.\ a theory
that includes nondeterministic choice operators which distribute over
sequential composition.\SG{composition distributes over choice!} There are two versions of this phenomenon,
with and without blocking (i.e.\ the empty set); for economy of
presentation we concentrate on the case with blocking. We start out
with a few notions concerned with blocking.
\begin{defn}[Bounded theory]\label{def:point}
  We call a large Lawvere theory $L$ \emph{bounded} if $|L(0,1)|=1$.
\end{defn}
\noindent (The term \emph{bounded} is motivated by the fact that
$\bot$ is the bottom of a natural \emph{approximation ordering}
introduced later.) Boundedness is connected to tensoring, as follows.
\begin{defn}\label{def:lbot}
  We denote by $L_\bot$ the large Lawvere theory generated by a
  constant $\bot$ and no equations.
\end{defn}
\begin{lem}\label{lem:bounded}
  A large Lawvere theory $L$ is bounded iff $L\tensor L_\bot\cong L$.
\end{lem}
\noindent We denote the only constant of a bounded theory by
$\bot_{0,1}$, and put $\bot_{n,m}=(\bot_{0,1}\tensor m)[\sharp]:n\to
m$ for all $n,m$, where $\sharp$ is the unique map $0\to n$. In the
sequel, we mostly write $\bot$ in place of $\bot_{n,m}$. As usual, we
have a corresponding notion of \emph{bounded monad.}

\begin{expl}[Bounded effects]\label{ex:point}
  Besides $L_\bot$, basic examples of bounded theories include all
  forms of nondeterminism with blocking. Similarly, the list monad is
  bounded ($\bot$ is the empty list). By Lemma~\ref{lem:bounded}, the
  state monad transformer, being defined by tensoring, preserves
  boundedness; e.g.\ the partial state monad $S\to
  (S\times\argument)_\bot$ and the non-deterministic state monad $S\to
  \PSet(S\times\argument)$ are bounded.
\end{expl}
\noindent 
\begin{defn}[Additive theories]~\cite{Goncharov10} A large Lawvere
  theory $L$ is \emph{finitely additive} if $L$ is enriched over join
  semilattices, and \emph{completely additive} if $L$ is enriched over
  complete join semilattices (with $\bot$).
\end{defn}
\noindent Again, corresponding notions for monads are implied. Joins
serve to model nondeterministic choice. Recall that \emph{enrichment}
amounts to hom-sets carrying the structure in question, and
composition preserving it in both arguments; i.e.\ composition
distributes over choice and deadlock on both sides. The enrichment is
a property rather than part of the data constituting a (completely or
finitely) additive Lawvere theory, as by Lemma~\ref{lem:add-order}
proved later, the ordering is uniquely determined by the algebraic
structure. The relation of additive theories to tensors is the
following.
\begin{lem}\label{lem:additive}
  For a large Lawvere theory $L$, the following are equivalent.
  \begin{sparrmenumerate}
    \item $L$ is completely additive.
    \item $L\cong L\tensor L_\PSet$.
    \item $L$ is bounded and has a family of morphisms $\join_n:n\to
      1$, where $n$ ranges over all sets, such that for any surjection
      $\sigma:m\to n$,
    \begin{equation*}
      \join_n=\join_m[\sigma],
    \end{equation*}
    and for every $L$-morphism $f:m\to 1$,    \begin{equation*}
      f\comp(\join_n\tensor m)=\join_n\comp (n\tensor f).
    \end{equation*} 
  \end{sparrmenumerate}
  The same equivalence holds for finite additivity, tensoring with
  $L_{\PSet_\omega}$, and (iii) for finite $n,m$.
\end{lem}
\noindent The operations $\join_n$ are $n$-fold joins, with $0$-fold
join $\join_0$ necessarily being $\bot$. In other words, a completely
additive theory $L$ is one that has nondeterministic choice operators
that commute over all operations of $L$ as prescribed by the tensor
law. From the above, it is immediate that
\begin{quote}
  \emph{tensoring a large Lawvere theory $L$ with $L_\PSet$ yields the
    free completely additive theory over $L$,}
\end{quote}
i.e.\ the (overlarge) category of completely additive theories is
reflexive in the category of large Lawvere theories.

\begin{expl}
  The generic example of a completely additive monad is $\PSet$, with
  joins being set unions. More generally, nondeterministic global
  state, $S\to\PSet(S\times\argument)$, is completely additive. A
  non-example is non-determinism with exceptions,
  $\PSet(\argument+E)$, which has several constants and hence fails to
  be bounded.
\end{expl}
\begin{rem}
  Although completely additive theories are enriched, they can be
  treated as standard large Lawvere theories --- as made explicit in
  Lemma~\ref{lem:additive}, the completely additive structure is
  algebraic (although unranked), and hence respected by all
  product-preserving functors. 
\end{rem}
%
%
\noindent
We proceed to formalize the example application from the introduction,
i.e.\ to show that completely additive monads indeed allow for a
generic Fischer-Ladner encoding of control structures. We base this
formalization on the fact that every completely additive monad is a
\emph{Kleene monad}~\cite{Goncharov10}, i.e.\ supports Kleene
iteration. Specifically, we can extend the computational metalanguage
with operators $\bot$ (deadlock), $+$ (binary choice, interpreted by
binary joins), and a generic loop construct
\begin{equation*}
  \infrule
{\Gamma\rhd p:TA ~~
 \Gamma,x:A\rhd q:TA}
{\Gamma\rhd\starTerm{x\leteq p}{q}:TA}.
\end{equation*}
The latter is interpreted as the join of all finite iterations of $q$,
prefixed with $x\leteq p$ and with the result $x$ of the computation
fed through the loop; i.e.\ $\starTerm{x\leteq p}{q}$ is the join of
$p$ and all programs $\letTerm{x\leteq p;x\leteq q;\dots;x\leteq
  q}{q}$ where $x\leteq q$ appears $n\ge 0$ times. Moreover, we
include in the signature the \emph{test operator} $?:2\to T 1$, which
sends $\inr\unit$ to $\ret\unit$ and $\inl\unit$ to $\nil$. Of course,
$+$, $\bot$, and $?$ are supported already by finitely additive
monads. From $+$, we can define an ordering $\leq$ in the usual way
via $p\leq q\iff p+q=q$. Issues in the axiomatization of a language
with choice and iteration are studied
in~\cite{GoncharovSchroderAtAl09,Goncharov10}. Relevant axioms and
rules for choice, deadlock, and iteration, including in particular two
induction rules for iteration, are shown in Figure~\ref{fig:kleene}.

\begin{figure*} 
\fbox{\parbox{\myboxwidth}{
\vspace{1ex}
\begin{equation*}
    \renewcommand{\arraystretch}{1.4}
  \begin{array}{lrl@{\qquad\qquad}lrl}
 \textbf{(plus$\nil$)} & p+ \nil & = p & 
\textbf{(comm)} & p+ q & = q + p\\
 \textbf{(idem)} &
p+ p & = p&
\textbf{(assoc)}&p+(q+ r) & = (p+ q)+ r\\
 \textbf{(bind$\nil_1$)}&
\letTerm{x\leteq p}{\nil} & = \nil&
\textbf{(bind$\nil_2$)} &\letTerm{x\leteq\nil}{p}&= \nil\\
 \textbf{(distr$_1$)}	&\qquad\letTerm{x\leteq p}{(q+ r)} & 
\multicolumn{3}{l}{= \letTerm{x\leteq p}{q}+\letTerm{x\leteq p}{r}}\\
 \textbf{(distr$_2$)} 	&\letTerm{x\leteq (p+ q)}{r} &
\multicolumn{3}{l}{= \letTerm{x\leteq p}{r}+\letTerm{x\leteq q}{r}}
\end{array}
\end{equation*}
~~\dotfill~~~
\begin{align*}
&\textbf{(unf$_1$)}   &\starTerm{x\leteq p}{q} &= p+\letTerm{x\leteq(\starTerm{x\leteq p}{q})}{q}&\\
&\textbf{(unf$_2$)}   &\starTerm{x\leteq p}{q} &= p+\starTerm{x\leteq (\letTerm{x\leteq p;q})}{q}&\\
&\textbf{(init)}      &\starTerm{x\leteq (\letTerm{y\leteq p}{q})}{r} &= \letTerm{y\leteq p}{\starTerm{x\leteq q}{r}}\quad(y\notin FV(r))&
\end{align*}

\vspace{-0.3em}
\begin{displaymath}\displaystyle
 \lrule{ind$_1$}{\letTerm{x\leteq p}{q}\leq p}{\starTerm{x\leteq p}{q}\leq p}
 \qquad\quad
\lrule{ind$_2$}{\letTerm{x\leteq q}{r}\leq r}
               {\letTerm{x\leteq(\starTerm{x\leteq p}{q})}{r}\leq\letTerm{x\leteq p}{r}}
\end{displaymath}
}}
\caption{Axioms and rules for nondeterminism (top) and Kleene iteration
  (bottom)}
  \label{fig:kleene}
\vspace{-1em}
\end{figure*}

Recall that we have given a definition of the if-operator in terms of
the case operator in the base language (Section~\ref{sec:lawvere}), with
$2=1+1$ representing the Booleans.
%
The proof of the following result appeals only to right distributivity
of sequencing over choice and deadlock; we state it in less generality
here only for the sake of brevity.
\begin{prop}[Generic Fischer-Ladner encoding]\label{prop:fl_dec}
~~~
  \begin{sparenumerate}
  \item Given a finitely additive monad $T$, for all $\Gamma\rhd b:2$, $\Gamma\rhd p: T A$ and $\Gamma\rhd q:T A$,
    \begin{displaymath}
      \ifTerm{b}{p}{q} = \letTerm{b?}{p}+\letTerm{(\neg b)?}{q}.
    \end{displaymath}
  \item Given a Kleene monad $T$, for every $\Gamma,x:A\rhd b: 2$ and
    $\Gamma, x:A\rhd p: T A$ the map sending $\Gamma,x:A\rhd q$ to
    \begin{displaymath}
      \Gamma,x:A\rhd\ifTerm{b}{\letTerm{x\leteq p}{q}}{\ret x}
    \end{displaymath}
    has a least fixed point given as the term
    \begin{equation*}
      \letTerm{x\leteq\bigl(\starTerm{x\leteq\ret
          x}{(\letTerm{b?}{p})}\bigr);(\neg b)?}{\ret x}.
    \end{equation*}
  \end{sparenumerate}
\end{prop}
\noindent The least fixed point in 2) is essentially a while loop,
which could be denoted in the form $\mathsf{while}\ b\ \mathsf{do}\
x\leteq p$. Here, the results of $p$ are fed through the loop via the
variable $x$; note that the condition $b$ itself does not read the
state (being of type $2$ instead of $T2$), but is updated in every
iteration of the loop via its dependence on the loop variable $x$.

\begin{rem}\label{rem:io}
  When absolutely free theories are used as a model of I/O,
  preservation of deadlock by sequencing from the left, as imposed by
  tensoring with $\PSet$, is hard to justify computationally. A
  satisfactory way of combining I/O with non-determinism and blocking
  will require fine-tuned mechanisms between sum and tensor yet to be
  developed.
\end{rem}

\section{Conservativity}
\label{sec:nondet-tensor}

\noindent As indicated above, the algebraic complexity of the tensor
implies that it is, in general, not at all clear that the component
theories map faithfully into the tensor, i.e.\ that adding a new
effect is \emph{conservative}. In the following we investigate this
issue for the powermonad obtained by tensoring with $\PSet$
(nondeterminism with blocking); in the terminology of the previous
section, this amounts to asking for which monads $T$ it can be soundly
assumed that they are completely additive, thus enabling, e.g.,
arguments using the Fischer-Ladner encoding.

Since $L_\PSet$ has a constant ($\emptyset$), an evident necessary
condition for $L\to L\tensor L_\PSet$ to be faithful is that $L$ can
conservatively be made bounded, i.e.\ $L\to L\tensor L_\bot$
(Definition~\ref{def:lbot}) must be faithful --- this is equivalent to
$L$ being already bounded if $L$ has a constant, and a complex issue
not in scope of the current investigation otherwise (one sufficient
condition is that $L$ is generated by equations having the same free
variables on both sides). For the sake of readability, we thus
restrict the further development to bounded large Lawvere theories.

However, constants are not the only problem: Even tensoring with
nonempty powerset $\PSetN$ can be non-conservative, one counterexample
being $(\PSetN)^2\tensor\PSetN=\PSetN$ where $(\PSetN)^2=\PSetN\PSetN$
is the double nonempty powerset monad (which may be thought of as
generated by unbounded conjunction and disjunction operators and a
distributive law). Collapse of $(\PSetN)^2\tensor\PSetN$ to $\PSetN$
is due to a variant of the well-known Eckmann-Hilton
argument~\cite{EckmannHilton62}. 

%
We proceed to give an exact characterization of those theories $L$ for
which tensoring with nondeterminism $L_\PSet$ is order-theoretically
conservative in a sense to be made precise presently. The main point
here is that bounded large Lawvere theories carry a natural
preordering:
\begin{defn}[Approximation]
  Let $L$ be a bounded large Lawvere theory. We compare elements of hom-sets
  $L(n,m)$ under the \emph{approximation preorder} $\sqsubseteq$,
  which is the smallest preorder (strictly speaking: family of
  preorders on hom-sets) with~$\bot$ as a bottom element and closed
  under the rule
  \begin{equation*}
    \lrule{$\pi_{\sqsubseteq}$}{\forall i.\,[\kappa_i]\comp f\sqsubseteq[\kappa_i]\comp g}
    {hf\sqsubseteq hg}
\end{equation*}
(equivalently, tupling and composition are monotone). 
\end{defn}
\noindent Roughly, $f\sqsubseteq g$ if $f$ is obtained from $g$ by
repeatedly deleting subterms and applying the given equations. It is
precisely the approximation preorder which provides the essential
handle for characterizing conservativity. The approximation preorder
does rely essentially on boundedness, i.e.\ on the fact that there is
exactly one constant. To find a corresponding result for tensoring
with non-empty non-determinism ($\PSet^\star$), i.e.\ to find a
replacement for the approximation preorder in the absence of $\bot$,
remains an open problem.
\begin{expl}
  \begin{sparenumerate}
  \item The approximation ordering on $L_\PSet$ and its variants is
    the subset relation. More generally, the approximation ordering
    coincides with the induced ordering in any additive theory, see
    Lemma~\ref{lem:add-order} below.
  \item In the list monad, $l\sqsubseteq k$ for lists $l,k$ iff $l$
    can be obtained from $k$ by deleting some of its entries.
  \item The approximation ordering on the theory $L_{\mathit{mult}}$
    of weighted nondeterminism is multiset containment.
  \item The approximation ordering on the partial state monad
    $S\to(S\times\argument)_\bot$ is the extension ordering.
  \end{sparenumerate}
\end{expl}
\begin{lem}\label{lem:add-order}
  Let $L$ be a finitely additive (hence bounded) large Lawvere theory. Then
  the approximation preorder $\sqsubseteq$ on $L$ coincides with
  the order $\le$ induced by the additive structure.
\end{lem}
\begin{lem}
  Every morphism of bounded large Lawvere theories preserves the
  approximation preorder.
\end{lem}
\noindent The previous lemma applies in particular to the tensor map
$L\to L\tensor L_\PSet$ for a bounded large Lawvere theory $L$. All
this indicates that the relevant notion of conservativity should take
into account the approximation preorder.
\begin{defn}\label{def:admit-unbounded}
  Let $L$ be a bounded large Lawvere theory. We say that $L$
  \emph{admits unbounded nondeterminism} if the tensor injection
  $\sigma_1:L\to L\tensor L_\PSet$ is an order embedding, i.e.\
  $\sigma_1$ is faithful and \emph{reflects the approximation
    ordering} in the sense that $f\sqsubseteq g$ whenever
  $\sigma_1(f)\le\sigma_1(g)$.
\end{defn}
\noindent That is, $L$ admits unbounded nondeterminism if tensoring
$L$ with $L_\PSet$ is \emph{order-theoretically conservative}.

For the remainder of this section, let $L$ be a bounded large Lawvere
theory. In a first step, we apply two key simplifications to the
description of $L\tensor L_\PSet$ given by the representation
according to \eqref{eq:seq_norm_body} (Section~\ref{sec:tensors}):
\eqref{eq:seq_norm_body} and the subsequent reduction imply that we
can represent a morphism $n\to m$ in $L\tensor L_\PSet$ as an
$m$-tuple of sets of $L$-morphisms $n+\{\emptyset\}\to 1$ (which may
be thought of as $L$-terms over $n+\{\emptyset\}$). We can, however
\begin{sparrmenumerate}
\item represent tuples of sets by sets of tuples using Cartesian
  products of sets, and
\item get rid of occurrences of $\emptyset$ in the bottom layer by
  replacing them with $\bot$, as $\bot=\emptyset$ in the tensor.
\end{sparrmenumerate}
Based on these observations, we arrive at a construction of the tensor
$L\tensor L_\PSet$ that can be proved correct independently of
Theorem~\ref{thm:contr}. To begin, we define a precursor of $L\tensor
L_\PSet$, a category $T_0$ whose objects are sets and whose morphisms
$n\to m$ are subsets of $L(n,m)$, with composition being complex
multiplication $AB=\{fg\mid f\in A,g\in B\}$, and identities
$\{\id\}$. We have identity-on-objects functors $\sigma_1^0:L\to T_0$
and $\sigma_2^0:L_{\PSet}\to T_0$ defined by $\sigma_1^0f=\{f\}$, and
by $\sigma_2^0(A_i)=\{[e]\mid e:m\to n,e(i)\in A_i\text{ for all
  $i$}\}$ for a morphism $(A_i):n\to m$ in $L_\PSet$, i.e.\ a family
of $m$ subsets $A_i\subseteq n$. The category $T_0$ inherits a functor
$I^0:\Set^\op\to T_0$ from $L$ via $\sigma_1^0$; it is easy to see
that under the axiom of choice (!), $I^0$ maps products to weak
products.

We then define a relation $\poweq$ on the sets $T_0(n,m)$ (strictly
speaking: a family of relations on hom-sets) inductively as the
smallest equivalence that contains all instances of the axiom scheme
\begin{equation*}
  (\bot)\quad\{\bot_{n,m}\}\poweq\emptyset\subseteq L(n,m)
\end{equation*}
and moreover forces uniqueness of tupling morphisms, i.e.\ is closed
under the infinitary rule
\begin{equation*}
  \lrule{$\pi$}{\forall i.\,[\kappa_i] A\approx[\kappa_i] B}
  {C A\approx C B}
\end{equation*}
where $L$-morphisms (such as $[\kappa_i]$) are meant to convert to
singletons when appropriate. We refer to $\poweq$ as \emph{rectangular
  equivalence}. Implied properties of $\poweq$ are symmetry and
congruence, the latter holding in particular for tupling and set
union. We put $T=T_0/{\poweq}$, and obtain functors $I:\Set^\op\to T$,
$\sigma_1:L\to T$, $\sigma_2:L_\PSet\to T$ by prolongation along
$T_0\to T$.
\begin{thm}\label{thm:p_char}
  The category $T$ of sets of $L$-morphisms modulo rectangular
  equivalence as constructed above is the tensor product $L\tensor
  L_{\PSet}$ of the bounded theory $L$ with unbounded nondeterminism
  $L_{\PSet}$.
\end{thm}
\noindent Similar results hold for tensoring with $L_{\PSetN}$ (in
fact, the construction for $L_\PSetN$ is slightly simpler) and for
tensoring finitary theories with $L_{\PSet_\omega}$ or $L_{\PSetNO}$.
Salient points in the proof are that the tensor law holds in $T_0$
\emph{up to rectangular equivalence}, and moreover that the general
tensor law justifies pointwise composition.

In $L$, we have morphisms $\Delta_i=\prod_j\delta_{ij}:n\to n$, where
for $i,j\in n$, $\delta_{ij}:1\to 1$ equals $[\id]$ if $i=j$ and
$\bot$ otherwise.
\begin{lem}\label{lem:delta}
  For $f:n\to m$, $g:m\to k$ in $L$, 
  \begin{math}
    fg\poweq\{f\Delta_i g\mid i\in m\}.
  \end{math}
\end{lem}
\noindent Since the right hand side of the above equivalence is a join
in the tensor $L\tensor L_\PSet$, order-theoretic conservativity will
imply that it is a join already in $L$. We proceed to develop a
characterization of the tensor in terms of order-theoretic closures
from this observation.

\begin{defn}
  We say that $A\subseteq L(n,m)$ is \emph{closed} if $A$ is
  downclosed and closed under the rule
  \begin{align*}
    \lrule{$\Delta$}{\forall i.\;g\Delta_i
      h\in A}{gh\in A}.
  \end{align*}
  We denote the smallest closed set containing $A\subseteq L(n,m)$ by
  $\downup{A}$. We write $\downup{f}$ for $\downup{\{f\}}$.
\end{defn}

\noindent The closure $\mathsf{cl}$ completely characterizes equality in
the tensor:

\begin{lem}\label{lem:p1_approx}
  For $A,B\subseteq L(n,m)$, $A\poweq B$ iff $\downup{A}=\downup{B}$.
\end{lem}
\noindent To prove this core fact, we need a preliminary lemma:
\begin{lem}\label{lem:downup-cong}
  Let $A:n\to m$ in $T_0$. Then for all $a:n\to m$ and all $b:m\to k$,
  \begin{equation*}
    a\in\downup{A}\implies ba\in\downup{bA}.
  \end{equation*}
\end{lem}
\begin{IEEEproof}
  Show that the set $\{a\in\downup{A}\mid ba\in\downup{bA}\}$
  contains $A$, is downward closed, and is closed under
  ($\Delta$). 
\end{IEEEproof}

\begin{IEEEproofNoQED}[Proof of Lemma~\ref{lem:p1_approx}]
  \emph{Only if:} Show that the equivalence $\simeq$ defined by
  $A\simeq B$ iff $\downup{A}=\downup{B}$ is closed under $(\pi)$ and
  contains all instances of $(\bot)$. Here, left congruence can
  conveniently be split off from $(\pi)$ as a separate rule, and
  closedness under left congruence is proved using
  Lemma~\ref{lem:downup-cong}.

  \emph{If:} It suffices to show that for $A:n\to m$ in $T_0$,
  $A\poweq\downup{A}$. Since $\poweq$ is congruent w.r.t.\ set
  union, it suffices to show that $A\poweq A\cup\{f\}$ for all
  $f\in\downup{A}$, which will follow if we show that the set 
  \begin{equation*}
    \bar A:=\{f:n\to m\mid A\poweq A\cup\{f\}\}
  \end{equation*}
  (which clearly contains $A$) is downward closed and closed under $\Delta$.
  \begin{sparitemize}
  \item \emph{$\bar A$ is downward closed:} define a preorder $\unlhd$
    by $f\unlhd g:\iff\{f,g\}\poweq g$; then $\unlhd$ is easily
    seen to be closed under $\pi_{\sqsubseteq}$, and hence contains
    $\sqsubseteq$. Now let $g\in\bar A$, $f\sqsubseteq g$. Then $f\unlhd g$ and therefore
    \begin{equation*}
      A\cup\{f\}\poweq A\cup\{g\}\cup\{f\}\poweq A\cup\{g\}\poweq A,
    \end{equation*}
    using congruence of $\poweq$ w.r.t.\ union.
  \item \emph{$\bar A$ is closed under ($\Delta$):} Let $f:n\to m$,
    $g:m\to k$, and let $f\Delta_i g\in\bar A$ for all $i\in m$. Then
    \begin{equation*}
      A\cup\{fg\}\poweq A\cup\{f\Delta_i g\mid i\in m\}
      \poweq A,
    \end{equation*}
    using congruence w.r.t.\ union and Lemma~\ref{lem:delta}. \myqed
  \end{sparitemize}
\end{IEEEproofNoQED}

\noindent Consequently, \emph{the tensor $L\tensor L_\PSet$ can be
  regarded as having closed subsets of $L(n,m)$ as morphisms $n\to
  m$}. The following is, then, more or less immediate.
%
\begin{thm}[Order-theoretic conservativity]\label{thm:ord-cons}
  Let $L$ be a bounded large Lawvere theory with approximation
  preorder $\sqsubseteq$ as defined above, and let $\sigma_1:L\to
  L\tensor L_\PSet$ be the tensor injection into the powermonad.
  \begin{sparenumerate}
  \item \label{item:reflect-eq}The following are equivalent:
    \begin{sparrmenumerateii}
    \item $\sigma_1$ reflects the approximation ordering.
    \item For all $f:n\to m$, $g:m\to k$ in $L$, $fg$ is a least
      upper bound of $\{f\Delta_ig\mid i\in m\}$.
    \item For all $f:n\to m$ in $L$, $\downup{f}=\{g\mid g\sqsubseteq
      f\}$.
    \end{sparrmenumerateii}
  \item If the equivalent conditions of \ref{item:reflect-eq}) are
    satisfied, then $\sigma_1$ is monic iff $\sqsubseteq$ is a
    partial order.
  \end{sparenumerate}
\end{thm}
\noindent Summarizing the above, \emph{a bounded large Lawvere theory
  $L$ admits unbounded nondeterminism iff the approximation
  preorder on $L$ is a partial order and for all $f:n\to m$,
  $g:m\to k$ in $L$,}
\begin{equation*}\textstyle
  \label{eq:cont}
  fg=\bigsqcup_{i\in m} f\Delta_ig.
\end{equation*}
Suprema of this form are preserved in the tensor.
\begin{rem}
  There is a variant of Theorem~\ref{thm:ord-cons} for tensoring
  finitary Lawvere theories with finite non-determinism, in which
  $n,m,k$  are finite in (ii), (iii).
\end{rem}
\begin{rem}[Equational conservativity]\label{rem:eq-cons}   
  One may wonder whether restricting to equational logic leads to
  weaker conditions for conservativity, which in the equational
  setting will be understood as faithfulness of the tensor map
  $\sigma_1:L\to L\tensor L_{\PSet}$. However, the conditions for
  order-theoretic conservativity of Theorem~\ref{thm:ord-cons} turn
  out to be necessary already for faithfulness of $\sigma_1$ under the
  mild additional assumption that $L$ is \emph{simply ordered}, i.e.\
  given any upper bound $h$ of $A\subseteq L(n,m)$, there exist $f$,
  $g$ such that $f g$ is an upper bound of $A$, $f g\sqsubseteq h$,
  and for every $i$ there is $a\in A$ for which $f\Delta_i
  g\sqsubseteq a$. All example theories mentioned so far are simply
  ordered. For simply ordered theories, closed sets are closed under
  all existing suprema, similarly to Scott closed sets, and thus all
  existing suprema are preserved in the powermonad.
\end{rem}
\begin{expl}\label{expl:cons}
  All absolutely free theories $L$ with at most one constant, such as
  input and output, map faithfully into $L\tensor L_\bot$, which can
  be shown to admit unbounded nondeterminism by
  Theorem~\ref{thm:ord-cons} (see however
  Remark~\ref{rem:io}). 

  The partial state monad $S\to (S\times\argument)_\bot$ admits
  unbounded nondeterminism, and Lemma~\ref{lem:p1_approx} allows
  identifying the tensor as the nondeterministic state monad $S\to
  \PSet(S\times\argument)$ (this can also be obtained from the known
  description of tensors with the state
  monad~\cite{HylandPlotkinAtAl06}). 

  Every finitely additive finitary Lawvere theory admits unbounded
  nondeterminism.  Hence, \emph{adding finite nondeterminism to a
    finitary theory is conservative iff adding unbounded
    nondeterminism is conservative}.  

  Multisets do not admit nondeterminism: the upper bound of
  $\{a,\bot\}$ and $\{\bot,a\}$ is not $\{a,a\}$ but
  $\{a\}$. Similarly, lists do not admit unbounded nondeterminism, as
  $[a,b]$ is not a supremum of $[a,\bot]=[a]$ and $[\bot,b]=[b]$. In
  both cases, already faithfulness of the tensor map fails.
\end{expl}

%

\section{Conclusion}

\noindent We have proved the existence of tensors of large Lawvere
theories for the case that one of the components is uniform. This
implies in particular that one can always tensor with unbounded
nondeterminism and with continuations, in the latter case improving a
previous existence result~\cite{HylandLevyAtAl07}. We have then given
a characterization of bounded theories that can be conservatively
tensored with nondeterminism, which means precisely that one can
assume such theories to be \emph{completely additive}. Completely
additive theories support a calculus for Kleene iteration, in
generalization of classical Kleene algebra, and, e.g., admit a
generalized form of the classical Fischer-Ladner
encoding~\cite{FischerLadner79}.

Neither the present work nor~\cite{GoncharovSchroder11} cover tensors
with finite powerset, whose existence remains an open question.
Although our results already have a quite order-theoretic flavour, an
important issue for future research is whether similar results can be
obtained in a domain-theoretic setting, using cpo-enriched Lawvere
theories. Another direction for extending our results is to generalize
them to enrichment over a topos, with a view to covering
presheaf-based effects such as local state~\cite{Power06} or name
creation~\cite{Stark08}.


\subsubsection*{Acknowledgments}
\noindent We gratefully acknowledge useful insights of various
contributors shared through the categories mailing list, in particular
Gordon Plotkin, Peter Johnstone, and Paul Levy.  Erwin R.\ Catesbeiana
has commented on inconsistent Lawvere theories.




%



\bibliographystyle{IEEEtran}
\bibliography{monads}

 \providecommand{\CoFI}{CoFI} \providecommand{\CASL}{CASL}
  \providecommand{\HasCASL}{HasCASL}
  \providecommand{\cofiWWW}{http://www.brics.dk/Projects/CoFI}
  \providecommand{\cofiFTP}{ftp://ftp.brics.dk/Projects/CoFI}
  \providecommand{\URL}[1]{#1} \providecommand{\footlink}[2]{#1\footnote{#2}}
  \providecommand{\cofiSN}[1]{Study Note~#1}
  \providecommand{\cofiLDSN}[1]{Language Design Study Note~#1}
  \providecommand{\cofiLDSNnoWWW}[1]{Language Design Study Note~#1}
  \providecommand{\cofiNote}[1]{Note~#1}
  \providecommand{\cofiDissentNote}[1]{Note~#1}
  \providecommand{\cofiDocument}[1]{Documents/#1}
  \providecommand{\cofiTentativeDocument}[1]{Documents/Tentative/#1}
  \providecommand{\cofiFTPlink}[2]{FTP:~#2}  \providecommand{\CoFI}{CoFI}
  \providecommand{\CASL}{CASL} \providecommand{\HasCASL}{HasCASL}
  \providecommand{\cofiWWW}{http://www.brics.dk/Projects/CoFI}
  \providecommand{\cofiFTP}{ftp://ftp.brics.dk/Projects/CoFI}
  \providecommand{\URL}[1]{#1} \providecommand{\footlink}[2]{#1\footnote{#2}}
  \providecommand{\cofiSN}[1]{Study Note~#1}
  \providecommand{\cofiLDSN}[1]{Language Design Study Note~#1}
  \providecommand{\cofiLDSNnoWWW}[1]{Language Design Study Note~#1}
  \providecommand{\cofiNote}[1]{Note~#1}
  \providecommand{\cofiDissentNote}[1]{Note~#1}
  \providecommand{\cofiDocument}[1]{Documents/#1}
  \providecommand{\cofiTentativeDocument}[1]{Documents/Tentative/#1}
  \providecommand{\cofiFTPlink}[2]{FTP:~#2}
\begin{thebibliography}{10}
\providecommand{\url}[1]{#1}
\csname url@samestyle\endcsname
\providecommand{\newblock}{\relax}
\providecommand{\bibinfo}[2]{#2}
\providecommand{\BIBentrySTDinterwordspacing}{\spaceskip=0pt\relax}
\providecommand{\BIBentryALTinterwordstretchfactor}{4}
\providecommand{\BIBentryALTinterwordspacing}{\spaceskip=\fontdimen2\font plus
\BIBentryALTinterwordstretchfactor\fontdimen3\font minus
  \fontdimen4\font\relax}
\providecommand{\BIBforeignlanguage}[2]{{%
\expandafter\ifx\csname l@#1\endcsname\relax
\typeout{** WARNING: IEEEtran.bst: No hyphenation pattern has been}%
\typeout{** loaded for the language `#1'. Using the pattern for}%
\typeout{** the default language instead.}%
\else
\language=\csname l@#1\endcsname
\fi
#2}}
\providecommand{\BIBdecl}{\relax}
\BIBdecl

\bibitem{Moggi91}
E.~Moggi, ``Notions of computation and monads,'' \emph{Inf.\ Comput.}, vol.~93,
  pp. 55--92, 1991.

\bibitem{JacobsPoll03}
B.~Jacobs and E.~Poll, ``{Coalgebras and Monads in the Semantics of Java},''
  \emph{Theoret.\ Comput.\ Sci.}, vol. 291, pp. 329--349, 2003.

\bibitem{Papaspyrou01}
N.~Papaspyrou, ``Denotational semantics of {ANSI} {C},'' \emph{Computer
  Standards and Interfaces}, vol.~23, pp. 169--185, 2001.

\bibitem{ShinwellPitts05}
M.~Shinwell and A.~Pitts, ``On a monadic semantics for freshness,''
  \emph{Theoret.\ Comput.\ Sci.}, vol. 342, pp. 28--55, 2005.

\bibitem{Harrison06}
W.~Harrison, ``The essence of multitasking,'' in \emph{Algebraic Methodology
  and Software Technology, {AMAST} 2006}, ser. LNCS, vol. 4019.\hskip 1em plus
  0.5em minus 0.4em\relax Springer, 2006, pp. 158--172.

\bibitem{Wadler97}
P.~Wadler, ``How to declare an imperative,'' \emph{ACM Comput.\ Surveys},
  vol.~29, pp. 240--263, 1997.

\bibitem{MoggiSabry01}
E.~Moggi and A.~Sabry, ``Monadic encapsulation of effects: A revised approach
  (extended version),'' \emph{J.\ Funct.\ Prog.}, vol.~11, pp. 591--627, 2001.

\bibitem{Lawvere63}
W.~Lawvere, ``Functorial semantics of algebraic theories,'' \emph{Proc.\ Natl.\
  Acad.\ Sci.\ USA}, vol.~50, no.~5, pp. 869--872, 1963.

\bibitem{PlotkinPower02}
G.~Plotkin and J.~Power, ``Notions of computation determine monads,'' in
  \emph{Foundations of Software Science and Computation Structures, FoSSaCS
  2002}, ser. LNCS, vol. 2303.\hskip 1em plus 0.5em minus 0.4em\relax Springer,
  2002, pp. 342--356.

\bibitem{CenciarelliMoggi93}
P.~Cenciarelli and E.~Moggi, ``A syntactic approach to modularity in
  denotational semantics,'' in \emph{Category Theory and Computer Science, CTCS
  1993}, 1993.

\bibitem{LiangHudakAtAl95}
S.~Liang, P.~Hudak, and M.~Jones, ``Monad transformers and modular
  interpreters,'' in \emph{Principles of Programming Languages, POPL 95}.\hskip
  1em plus 0.5em minus 0.4em\relax ACM Press, 1995, pp. 333--343.

\bibitem{HylandLevyAtAl07}
M.~Hyland, P.~Levy, G.~Plotkin, and J.~Power, ``Combining algebraic effects
  with continuations,'' \emph{Theoret.\ Comput.\ Sci.}, vol. 375, pp. 20 -- 40,
  2007.

\bibitem{HylandPlotkinAtAl06}
M.~Hyland, G.~Plotkin, and J.~Power, ``Combining effects: Sum and tensor,''
  \emph{Theoret.\ Comput.\ Sci.}, vol. 357, pp. 70--99, 2006.

\bibitem{PowerShkaravska04}
J.~Power and O.~Shkaravska, ``From comodels to coalgebras: State and arrays,''
  in \emph{Coalgebraic Methods in Computer Science, CMCS 2004}, ser. ENTCS,
  vol. 106.\hskip 1em plus 0.5em minus 0.4em\relax Elsevier, 2004, pp.
  297--314.

\bibitem{Stark08}
I.~Stark, ``Free-algebra models for the {$\pi$}-calculus,'' \emph{Theoret.\
  Comput.\ Sci.}, vol. 390, pp. 248--270, 2008.

\bibitem{TixKeimelAtAl09}
R.~Tix, K.~Keimel, and G.~Plotkin, ``Semantic domains for combining probability
  and non-determinism,'' \emph{ENTCS}, vol. 222, pp. 3--99, 2009.

\bibitem{VaraccaWinskel06}
D.~Varacca and G.~Winskel, ``Distributing probability over non-determinism,''
  \emph{Math.\ Struct.\ Comput.\ Sci.}, vol.~16, pp. 87--113, 2006.

\bibitem{GoncharovSchroder11}
S.~Goncharov and L.~Schr{\"o}der, ``A counterexample to tensorability of
  effects,'' DFKI, Tech. Rep., 2011.

\bibitem{FischerLadner79}
M.~Fischer and R.~Ladner, ``Propositional dynamic logic of regular programs,''
  \emph{J.\ Comput.\ Sys.\ Sci.}, 1979.

\bibitem{Plotkin77}
G.~Plotkin, ``{LCF} considered as a programming language,'' \emph{Theoret.\
  Comput.\ Sci.}, vol.~5, pp. 223--255, 1977.

\bibitem{Dubuc70}
E.~Dubuc, \emph{Kan Extensions in Enriched Category Theory}, ser. LNM.\hskip
  1em plus 0.5em minus 0.4em\relax Springer, 1970, vol. 145.

\bibitem{AbramskyJung94}
S.~Abramsky and A.~Jung, ``Domain theory,'' in \emph{Handbook of Logic in
  Computer Science}.\hskip 1em plus 0.5em minus 0.4em\relax Oxford University
  Press, 1994, vol.~3, pp. 1--168.

\bibitem{DrosteKuichAtAl09}
M.~Droste, W.~Kuich, and H.~Vogler, Eds., \emph{Handbook of Weighted
  Automata}.\hskip 1em plus 0.5em minus 0.4em\relax Springer, 2009.

\bibitem{Peyton-Jones03}
S.~Peyton-Jones, Ed., \emph{{Haskell} 98 Language and Libraries --- The Revised
  Report}.\hskip 1em plus 0.5em minus 0.4em\relax Cambridge University Press,
  2003, also: J.\ Funct.\ Prog.\ {{\bf 13}} (2003).

\bibitem{HylandPower07}
M.~Hyland and J.~Power, ``The category theoretic understanding of universal
  algebra: Lawvere theories and monads,'' \emph{ENTCS}, vol. 172, pp. 437--458,
  2007.

\bibitem{Goncharov10}
S.~Goncharov, ``Kleene monads,'' Ph.D. dissertation, Universit{\"a}t Bremen,
  2010.

\bibitem{GoncharovSchroderAtAl09}
S.~Goncharov, L.~Schr\"{o}der, and T.~Mossakowski, ``Kleene monads: handling
  iteration in a framework of generic effects,'' in \emph{Algebra and Coalgebra
  in Computer Science, CALCO 2009}, ser. LNCS, vol. 5728.\hskip 1em plus 0.5em
  minus 0.4em\relax Springer, 2009, pp. 18--33.

\bibitem{EckmannHilton62}
B.~Eckmann and P.~Hilton, ``Group-like structures in general categories {I} --
  multiplications and comultiplications,'' \emph{Math.\ Ann.}, vol. 145, pp.
  227--255, 1962.

\bibitem{Power06}
J.~Power, ``Semantics for local computational effects,'' in \emph{Mathematical
  Foundations of Programming Semantics, MFPS 2006}, ser. ENTCS, vol. 158, 2006,
  pp. 355--371.

\end{thebibliography}

\newpage
\appendix\allowdisplaybreaks

(In the appendix, we use the term Lawvere theory to mean large Lawvere
theory.)
\subsection{Proof of Theorem~\ref{thm:contr}.}
\noindent The proof is by explicit syntactic construction of the tensor product ${L=L_1\tensor L_2}$.
To begin, we define a (not necessarily locally small) category $\BC$
on top of $L_1$, $L_2$ as follows. For $f\in\Hom_{L_2}(k,m)$ and
$g\in\Hom_{L_1}(n,k)$ let $f\mix_k g$ be a synonym for the pair
$\brks{f, g}$. We agree to omit the subscript at $\mix$ if it is clear
from the context. We also agree that $\mix$ binds weaker than
composition. Let us define objects of $\BC$ to be sets and morphisms
from $\Hom_{\BC}(n,m)$ to be finite paths
\begin{displaymath}
\mypath{f_1\mix g_1\comma\ldots\comma f_k\mix g_k}
\end{displaymath}
adhering to the typing constraints: $n$ is the source of $g_k$, $m$ is
the target of $f_1$, and for $i=1,\dots,k-1$, the source of $g_{i+1}$
is the target of $f_i$. We often omit brackets for one-element
paths.

The identity morphisms of $\BC$ are the empty paths, and composition is concatenation of paths. Clearly, $\BC$ is a category. 
On every hom-set of $\BC$ we define an equivalence relation $\sim$ as the equivalence generated by the clauses
\begin{align*}
\mypath{\ldots\comma[\id]\mix [\id]\comma\ldots}\sim&\, \mypath{\ldots\comma\ldots},\\
\mypath{\ldots\comma f[e]\mix g\comma\ldots}\sim&\, \mypath{\ldots\comma f\mix [e] g\comma\ldots}\\
\end{align*}
and
\begin{multline*}
  \mypath{\ldots\comma f\mix (n'\tensor g)\comma (f'\tensor m)\mix g'\comma\ldots}\\\sim\,\mypath{\ldots\comma
    f (f'\tensor m')\mix (n\tensor g) g'\comma\ldots}
\end{multline*}
where $f':n\to n'$ and $g:m\to m'$. By construction, $\sim$ is a congruence on $\BC$, so that we have a quotient category $\BC/_{\sim}$. Using the fact that $L_2$ is uniform, we show that every morphism of $\BC/_{\sim}$ has a representative of the form
\begin{equation}\label{eq:seq_norm}
\mypath{f\mix g\comma\cn^n_{L_2}\mix[\id]}.
\end{equation}
%
To that end, let us take any morphism $f$ of $\BC$. By attaching sufficiently many elements $[\id]\mix[\id]$ in the end of $f$ we ensure that its length is at least $2$ and its last element is $[\id]\mix[\id]$. Then we successively apply the following reduction sequence, whose net effect is length-decreasing, as long as possible:
\begin{flalign*}
\mypath{f_1\mix &\,g_1\comma f_2\mix g_2\comma\ldots}\\
\sim\,&\mypath{f_1\mix g_1\comma (\hat f_2\tensor m_2) [u_{f_2}]\cn^{k_1}_{L_2}\mix g_2\comma\ldots}\\
=\,&\mypath{f_1\mix g_1\comma (\hat f_2\tensor m_2)([u_{f_2}]\cn^{k_1}_{L_2}\tensor 1)\\
  &\hspace{10em}\mix (k_1\tensor [\id]) g_2\comma\ldots}\\
\sim\,&\mypath{f_1\mix g_1\comma (\hat f_2\tensor m_2)\mix ((s\times m_2)\tensor [\id])\comma\\
&\hspace{8em}([u_{f_2}]\cn^{k_1}_{L_2}\tensor 1)\mix g_2\comma\ldots}\\
=\,&\mypath{f_1\mix g_1\comma (\hat f_2\tensor m_2)\mix[\id]\comma [u_{f_2}]\cn^{k_1}_{L_2}\mix g_2\comma\ldots}\\
\sim\,&\mypath{f_1\mix g_1\comma (\hat f_2\tensor m_2)\mix[u_{f_2}]\comma\cn^{k_1}_{L_2}\mix g_2\comma\ldots}\\
=\,&\mypath{f_1\mix (1\tensor g_1)\comma (\hat f_2\tensor m_2)\mix[u_{f_2}]\comma\cn^{k_1}_{L_2}\mix g_2\comma\ldots}\\
\sim\,&\mypath{f_1(\hat f_2\tensor k_1)\mix (s\tensor g_1) [u_{f_2}]\comma\cn^{k_1}_{L_2}\mix g_2\comma\ldots}.
\end{flalign*}
Here, $f_i:k_i\to m_i$, $g_i:n_i\to k_i$ and $\hat f_2:s\to 1$. At the last step of the reduction we obtain a pair of the form~\eqref{eq:seq_norm}.

Let us define $I:\Set^{op}\to\BC/_{\sim}$ by putting $I(n)=n$ for every set $n$ and $I(e)=[e]\mix [\id]$ for every set-function $e:n\to m$. 
We would like to prove that $I$ weakly preserves small products, i.e.\ the families $I(\kappa_i):\sum_i n_i\to n_i$ define weak small products\ednote{LS: This seems to be the only place where general products are used, and I think we can replace this by products of $1$} in~$\BC/_{\sim}$. Let $f_i:m\to n_i$ be a family of morphisms in $\BC/_{\sim}$. First we consider a special case when every $f_i$ is presentable by a one-element path, e.g.\ $f_i\sim g_i\mix_{k_i} h_i$. Since $L_2$ has all small products, there exists a morphism $h:m\to \sum_i k_i$ of $L_2$ such that for every $i$, $h_i=[\kappa_i] h$. Analogously, since $L_1$ has all small products, there exists a morphism $g:\sum_i k_i\to\sum_i n_i$ of $L_1$ such that for every $i$, $g_i [\kappa_i]=[\kappa_i] g$. The equality $f_i = I(\kappa_i)\mypath{g\mix h}$, characterizing weak products, now follows from the diagram
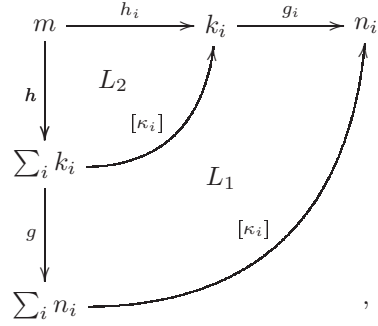
\begin{displaymath}
\xymatrix@R35pt@C40pt@M4pt{
m \ar[r]^{h_i} \ar[d]_{h}\ar@{}[dr]|>>>>>>>>>>>>>>>>{\hbox{$L_2$}} \ar[d]_{h}\ar@{}[ddrr]|>>>>>>>>>>>>>>>>>>>>>>>{\hbox{$L_1$}} & k_i \ar[r]^{g_i} & n_i\\
{\sum_i k_i} \ar[d]_{g} \ar@/_5ex/[ru]^{[\kappa_i]}     & &\\
{\sum_i n_i} \ar@/_9ex/[rruu]^{[\kappa_i]}& &,
}
\end{displaymath}
whose two cells are commuting in $L_1$ and in $L_2$ respectively. In general, the $f_i$ might not be presentable by one-element paths, but as we have argued above, they must be presentable by paths of length $2$. In particular, for every $i$, $f_i\sim g_i h_i$ where both $g_i:k_i\to n_i$ and $h_i:m\to k_i$ are one-element. As we have proved, there exists $h$ such that for every $i$, $h_i = I(\kappa_i) h$. On the other hand, every $g_i I(\kappa_i)$ is easily seen to be equivalent to a one-element path and therefore, there exists $g$ such that for every $i$, $g_i I(\kappa_i)=I(\kappa_i) g$. We have thus: $f_i\sim g_i h_i\sim g_i I(\kappa_i) h\sim I(\kappa_i) g h$ and we are done.

Let us prove that $\BC/_{\sim}$ is locally small. Since every morphism of $\BC$ has the form~\eqref{eq:seq_norm}, every hom-set of $\BC/_{\sim}$ has at most as many equivalence classes as  there are non-equivalent morphisms~\eqref{eq:seq_norm} in the corresponding hom-set of $\BC$. Let us fix some pair $\mypath{f\mix_k g\comma\cn^n_{L_2}\mix [\id]}\in\Hom_{\BC}(n,m)$. Let $c=\Hom_{L_2}(0,1)$. By local smallness of $L_2$, $s=\Hom_{L_2}(n+c,1)$ is a set, and thus $s\in\Ob(L_1)$. For every $i\in k$, $g_i=[\kappa_i] g$ belongs to $\Hom_{L_2}(n+c,1)$ and we denote by $u:k\to s$ the induced index transformation. Let $h:n+c\to s$ be the tupling morphism for the whole family $\Hom_{L_2}(n+c,1)$. Then
\begin{align*}
\mypath{f\mix_k g\comma\cn^n_{L_2}\mix [\id]}&\sim \mypath{f\mix_k [u] h\comma\cn^n_{L_2}\mix [\id]}\\&\sim \mypath{f [u]\mix_s h\comma\cn^n_{L_2}\mix [\id]}.
\end{align*}
We have thus shown that every morphism of $\Hom_{\BC/_{\sim}}(n,m)$ has a representative in the set $\Hom_{L_1}(m,s)\times\Hom_{L_2}(n,s)$ and hence $\Hom_{\BC/_{\sim}}(n,m)$ is also a set.  

%
%

Let $\approx$ be the smallest congruence on $\BC$, containing $\sim$ and closed under the rule:
\begin{equation}\label{eq:closure}
\forall i.\, [\kappa_i] f\approx [\kappa_i] g\implies f\approx g.
\end{equation}
We then have a canonical functor $\BC/_{\sim}\to\BC/_{\approx}$, which equips $\BC/_{\approx}$ with all small weak products, which due to~\eqref{eq:closure}  are in fact products. By postcomposing ${I:\Set^{op}\to \BC/_{\sim}}$ with the canonical projection, we obtain a product preserving functor ${I:\Set^{op}\to \BC/_{\approx}}$, so that $L=\BC/_{\approx}$ is a Lawvere theory. We will be done once we show that $L=L_1\tensor L_2$. 
We define functors $\sigma_i:L_i\to L$ by 
\begin{align*}
\sigma_1(f) = f\mix [\id], && \sigma_2(g) = [\id]\mix g
\end{align*} 
(omitting equivalence class formation from the notation).
The following calculation ensures commutativity of $\sigma_1$ and $\sigma_2$:
\begin{flalign*}
(\sigma_2(g)\,\tensor &\,n') (m\tensor\sigma_1(f))\\
=&\,(\mypath{[\id]\mix g}\tensor n') (m\tensor\mypath{f\mix [\id]})\\
=&\,\mypath{[\id]\mix g\tensor n'}\mypath{m\tensor f\mix [\id]}\\
=&\,[\id](m'\tensor f)\mix (g\tensor n) [\id]\\
=&\,(m'\tensor f)\mix (g\tensor n)\\
=&\,\mypath{m'\tensor f\mix [\id]}\mypath{[\id]\mix g\tensor n}\\
=&\,(m'\tensor\mypath{f\mix [\id]}) (\mypath{[\id]\mix g}\tensor n)\\
=&\,(m'\tensor\sigma_1(f)) (\sigma_2(g)\tensor n).
\end{flalign*}
Finally, let $L'$ be another Lawvere theory equipped with a pair of commuting morphisms $\alpha_i:L_i\to L'$. We define a morphism of categories $\alpha:\BC\to L'$ to be identity on objects and by the equations $\alpha\mypath{\,}=\,\id$ and
\begin{multline*}
\alpha\mypath{f_1\mix g_1\comma\ldots\comma f_k\mix g_k}=\\\alpha_1(f_1)\alpha_2(g_1)\ldots\alpha_1(f_k) \alpha_2(g_k)
\end{multline*}
on morphisms. It is straightforward to verify by definition that $f\approx g$ implies ${\alpha(f)=\alpha(g)}$. Therefore, by the characteristic property of the quotient category, $\alpha$ lifts to a morphism of Lawvere theories $\alpha:L\to L'$. It is again easy to verify that for $i=1,2$, $\alpha_i=\alpha\sigma_i$. Uniqueness of $\alpha$ is clear. Therefore $L$ is indeed a tensor product of $L_1$ and $L_2$ and we are done. \myqed

\subsection{Proof of Proposition~\ref{prop:fl_dec}}
  \begin{sparenumerate}
  \item Let for every $\Gamma\rhd t: T A\times T A$,
    $h(t)=\fst(t)+\snd(t)$. Then
    \begin{flalign*}
      \ifTerm{b}{\,&p}{q}\\
      =\,& \case{b}{\inl\unit\mapsto p}{\inr\unit\mapsto q}\\
      =\,& \case{b}{\inl\unit\mapsto h\brks{p,\nil}}{\inr\unit\mapsto h\brks{\nil,q}}\\
      =\,& h(\case{b}{\inl\unit\mapsto\brks{p,\nil}}{\inr\unit\mapsto\brks{\nil,q}})\\
      =\,& h(\case{b}{\inl\unit\mapsto p}{\inr\unit\mapsto\nil},\\
      &~~\case{b}{\inl\unit\mapsto\nil}{\inr\unit\mapsto q})\\
      =\,& \letTerm{(\case{b}{\inl\unit\mapsto\ret\unit}{\inr\unit\mapsto\nil})}{p}~+\\
      &\letTerm{(\case{b}{\inl\unit\mapsto\nil}{\inr\unit\mapsto\ret\unit})}{q}\\
      =\,& \letTerm{(\case{b}{\inl\unit\mapsto\ret\unit}{\inr\unit\mapsto\nil})}{p}~+\\
      &\letTerm{(\case{\neg b}{\inl\unit\mapsto\ret\unit}{\inr\unit\mapsto\nil})}{q}\\
      =\,&\letTerm{b?}{p}+\letTerm{(\neg b)?}{q}
    \end{flalign*}
    and we are done.
  \item First, note that by Lemma~\ref{lem:add-order}, $\sqsubseteq$
    coincides with $\leq$. By part (i), we need to show that
\begin{equation}\label{eq:while_def}
  \letTerm{x\leteq(\starTerm{x\leteq\ret x}{(\letTerm{b?}{p})});(\neg b)?}{\ret x}
\end{equation}
is the least fixed point of
\begin{equation}\label{eq:while_fp}
q~\mapsto~\letTerm{b?;x\leteq p}{q} + \letTerm{(\neg b)?}{\ret x}.
\end{equation}
First observe that~\eqref{eq:while_def} is a fixed point
of~\eqref{eq:while_fp}:
\begin{align*}
\letTerm{b?;\,&x\leteq p}{(\letTerm{x\leteq(\starTerm{x\leteq\ret x}{(\letTerm{b?}{p})});\\&(\neg b)?}{\ret x})} + \letTerm{(\neg b)?}{\ret x}\\
=\,&\letTerm{x\leteq(\letTerm{b?;x\leteq p}{\\&\starTerm{x\leteq\ret x}{(\letTerm{b?}{p})}}+\ret x);(\neg b)?}{\ret x}\\
=\,&\letTerm{x\leteq(\starTerm{x\leteq (\letTerm{b?}{p})}{(\letTerm{b?}{p})}\\&+\ret x);(\neg b)?}{\ret x}\\
=\,&\letTerm{x\leteq(\starTerm{x\leteq\ret x}{(\letTerm{b?}{p})});(\neg b)?}{\ret x}.
\end{align*}
In order to show that~\eqref{eq:while_def} is the least fixed point, suppose $q$ is some other fixed point of~\eqref{eq:while_fp}. Then
\begin{eqnarray*}
\letTerm{b?;x\leteq p}{q}\leq q\\
\letTerm{(\neg b)?}{\ret x}\leq q
\end{eqnarray*}
From the former inequality, by~\textbf{(ind$_2$)}:
\begin{displaymath}
\letTerm{x\leteq(\starTerm{x\leteq\ret x}{(\letTerm{b?}{p})})}{q}\leq q
\end{displaymath} 
from which we conclude by the latter inequality,
\begin{displaymath}
\letTerm{x\leteq(\starTerm{x\leteq\ret x}{(\letTerm{b?}{p})});(\neg b)?}{\ret x}\leq q.
\end{displaymath} 
Therefore~\eqref{eq:while_def} is indeed the least fixed point
of~\eqref{eq:while_fp} and the proof is thus completed.\myqed
  \end{sparenumerate} 

\subsection{Proof of Lemma~\ref{lem:add-order}}

\noindent To prove that $\sqsubseteq$ is contained in $\preceq$, it
suffices to show that $\preceq$ has the closure properties defining
$\sqsubseteq$. By definition, $\preceq$ has $\bot$ as bottom. To see
that $\preceq$ is closed under ($\pi_{\sqsubseteq}$), let $f,g:n\to m$
in $L$ such that $[\kappa_i]f\preceq[\kappa_j]g$ for all $i\in m$. By
definition, this means that
$[\kappa_i]f+[\kappa_j]g[\kappa_i](f+g)=[\kappa_j]g$ for all $i$, so
that $f+g=g$, i.e.\ $f\preceq g$; since by definition, composition is
monotone w.r.t.\ $\preceq$, it follows that $hf\preceq hg$ for $h:m\to
k$.

  To show that, conversely, $\preceq$ is contained in $\sqsubseteq$,
  let $f\preceq g$. Then $f=f+\bot\sqsubseteq f+g =g$.\myqed

\subsection{A Direct Construction of the Nonempty Powermonad}

\noindent To pave the ground for the direct construction of the
powermonad, i.e.\ the proof of Theorem~\ref{thm:p_char}, we describe
the direct construction of the nonempty powermonad, i.e.\ tensoring
with non-blocking unbounded nondeterminism.

We need a preliminary lemma to ease the proof of the tensor equation.
\begin{lem}
  In the notation of Definition~\ref{def:lawvere_tensor}, the tensor equation
  reduces to the case $m_1=m_2=1$. 
\end{lem}

\begin{IEEEproof}
  We prove the general case as follows: To check commutation of the
  requisite diagram for arbitrary $m_1,m_2$, it suffices to check
  commutation for all postcompositions with the product projections
  $\pi_{ij}=[(i,j)]:m_1\times m_2\to 1$ for $i\in m_1,j\in m_2$ (where
  $(i,j):1\to m_1\times m_2$ denotes the obvious constant map). Note that
  $\pi_{ij}=\pi_i (m_1\tensor \pi_j)=\pi_j(\pi_i\tensor m_2)$ where
  $\pi_i:m_1\to 1$ and $\pi_j:n_1\to 1$ are product projections. Therefore
  $\pi_{ij}(f_1\tensor m_2)=\pi_i(m_1\tensor \pi_j)(f_1\tensor
  m_2)=\pi_i(f_1\tensor\pi_j)=(\pi_if_1\tensor\pi_j)$ and $\pi_{ij}(m_1
  \tensor f_2)=\pi_j(\pi_i\tensor m_2)(m_1\tensor f_2)=\pi_j(\pi_i\tensor
  f_2)=\pi_i\tensor(\pi_jf_2)$. (Note here that for $f:n\to m$ and a map
  $e:k\to l$, $f\tensor[e]:n\times l\to m\times k$ is definable as the
  morphism into the $k$-fold product $m\times k$ whose postcomposition
  with the $j$-th projection $m\times k\to m$ ($j\in k$) is
  $f\pi_{e(j)}$, where $\pi_{e(j)}$ is the $e(j)$-th product
  projection $n\times l\to n$.)\ednote{LS: There's a number of
    slightly speculative properties being used here that deserve an
    extra lemma.} Next note that $(\pi_if_1\tensor \pi_j)(n_1\tensor
  f_2)=\pi_if_1(n_1\tensor\pi_jf_2)$ and $(\pi_i\tensor\pi_j f_2)(f_1\tensor
  n_2)=\pi_jf_2(\pi_if_1\tensor n_2)$, so that we are done by commutation
  of
  \begin{equation*}
    \xymatrix@R30pt@C30pt@M6pt{
      n_1\times n_2        \ar[r]^{n_1\tensor \pi_jf_2} \ar[d]_{\pi_i f_1\tensor n_2} & n_1\ar[d]^{\pi_if}\\
      n_2        \ar[r]^{\pi_jf_2} &1.
    }
  \end{equation*}
\end{IEEEproof}

Let $L$ be a Lawvere theory. We give a construction of the tensor
$T=L\tensor L_{\PSetN}$. We begin by constructing a category $T_0$
with an identity-on-objects functor $I_0:\Set\to T_0$, with the same
notation as for Lawvere theories, with the following properties:
\begin{sparitemize}
\item $I_0$ maps products to weak products;
\item $T_0$ has functors $F^0:L\to T_0$, $F^0_{\PSetN}:L_{\PSetN}\to
  T_0$ that commute with the respective functors from $\Set$
\end{sparitemize}
That is, $T_0$ will fail to be the tensor $L\tensor L_{\PSetN}$ on
two counts: tupling morphisms need not be unique in $T_0$, and the two
functors from the component theories into $T_0$ need not commute. One
of the surprises in the construction is that repairing the first
defect will remedy also the second one.

Morphisms $n\to m$ in $T_0$ are just nonempty sets of
$L$-morphisms. Composition is defined by $A B=\{a b\mid a\in
A,b\in B\}$; identities are singleton sets $\{\id_n\}$. The functor
$F^0:L\to T_0$ maps a morphism $f$ to the singleton $\{f\}$. We then
define $I_0$ as the composite $\Set\to L\to T_0$. The tupling
$\Brks{A_i}:n\to\sum m_i$ of $k$ $T_0$-morphisms $A_i:n\to m_i$ is
defined as
\begin{equation*}
  \Brks{A_i}=\{\Brks{f_i}\mid f_i\in A_i\text{ for all $i$}\},
\end{equation*}
where $\Brks{f_i}$ denotes tupling in $L$. We regard morphisms $n\to
m$ in $L_{\PSetN}$ as $m$-tuples $(A_i)$ of nonempty subsets of
$n$. Then the functor $F^0_{\PSetN}:L_{\PSetN}\to T_0$ maps $(A_i)$
to
\begin{equation*}
  F^0_{\PSetN}((A_i))=\{[e]\mid e:m\to n,e(i)\in A_i\text{ for all $i$}\}.
\end{equation*}
In the special case $m=1$, in which case a morphism $n\to m$ is just a
single subset $A\subseteq n$, note that
$F^0_{\PSetN}(A)=\{[\kappa_{i}]\mid i\in A\}$.

This completes the definition of $T_0$. We need to check a few properties:
\begin{sparitemize}
\item \emph{The composite $\Set\to L_{\PSetN}\to T_0$ coincides with
    $I_0$.} To see this, let $e:m\to n$ be a map. In $L_{\PSetN}$,
  $[e]:n\to m$ is the $m$-tuple $(\{e(i)\})_{i\in m}$. Under
  $F^0_{\PSetN}$, this becomes the set
  \begin{equation*}
    \{[\bar e]\mid \bar e:m\to n,\bar e(i)\in\{e(i)\}\text{ for all $i$}\}
    =\{[e]\}.
  \end{equation*}
\item \emph{The tupling morphisms project back to their components:}
  For $A_i:n\to m_i$ in $T_0$, we have
  \begin{multline*}\textstyle
    \pi_j\Brks{A_i}=\{\pi_j\Brks{f_i}\mid (f_i)\in\prod A_i
    \}=\\
    \textstyle\{f_j\mid (f_i)\in\prod A_i\}=A_j,
  \end{multline*}
  as the $A_i$ are nonempty (note that this requires the axiom of
  choice).
\end{sparitemize}

We now proceed to repair the mentioned defects by quotienting $T_0$ by
an appropriate equivalence relation. We define the relation $\approx$
as the smallest reflexive transitive relation closed under the
infinitary rule
\begin{equation*}
  \lrule{$\pi$}{\forall i.\,[\kappa_i]A\approx[\kappa_i]B}
  {CA\approx CB}.
\end{equation*}
\begin{lem}
  The relation $\approx$ satisfies the following properties.
  \begin{sparenumerate}
  \item $\approx$ is symmetric.
  \item $\approx$ is a congruence w.r.t.\ composition.
  \item $\approx$ is a congruence w.r.t.\ tupling.
  \end{sparenumerate}
\end{lem}
\begin{IEEEproofNoQED}
  \begin{sparenumerate}
  \item Put $\approx^s=\approx\cap\approx^-$, where $\cdot^-$ denotes
    the inverse relation. Clearly, $\approx^s$ is reflexive and
    transitive. Moreover, $\approx^s$ is easily seen to be closed
    under ($\pi$). Consequently, $\approx\subseteq\approx^s$, so that
    $\approx$ is symmetric.
  \item Let 
    \begin{equation*}
      A\approx^c B\iff \forall L,R.\,LAR\approx LBR.
    \end{equation*}
    Then $\approx^c$ is clearly reflexive and transitive. Moreover,
    $\approx^c$ is closed under ($\pi$): if
    $[\kappa_i]A\approx^c[\kappa_i]B$ for all $i$, then in particular
    $[\kappa_i]AR\approx[\kappa_i]BR$ for all $i,R$ and hence
    $LAR\approx LBR$ for all $L,R$. Therefore, $\approx^c$ contains
    $\approx$, so that $\approx$ is a congruence.
  \item Let $A_i,B_i:n\to m_i$ and $A_i\approx B_i$ for $i\in k$. To
    prove $\Brks{A_i}\approx\Brks{B_i}$, we have to show
    $[\kappa_{(i,j)}]\Brks{A_i}\approx[\kappa_{(i,j)}]\Brks{B_i}$ for
    $i\in k$, $j\in m_i$ (i.e.\ $(i,j)\in\sum_{i\in k}m_i$). Now
    $\kappa_{(i,j)}=(\kappa_j\times\id_k)\iota_i$, where
    $\iota_i:m_i\to\sum m_i$ is the coproduct embedding (and hence
    $[\iota_i]$ is a weak product projection). Thus, we have
    $[\kappa_{(i,j)}]\Brks{A_i}=[\kappa_j]A_i\approx[\kappa_j]B_i=
    [\kappa_{(i,j)}]\Brks{B_i}$, using the assumption $A_i\approx B_i$
    and the fact that $\approx$ is congruent w.r.t.\ composition.
  \end{sparenumerate}
\end{IEEEproofNoQED}
A last observation that needs to be made is that the tensor law holds
in $T_0$ modulo $\approx$: Let $A\subseteq m$, corresponding to the
morphism $\bar A=\{[\kappa_i]\mid i\in A\}:m\to 1$ in $T_0$, and let
$f:n\to 1$ in $L$ (identified with a singleton in $T_0$). We have to
show that the diagram
\begin{equation}
\begin{split}\label{eq:tensor}
\xymatrix@R30pt@C30pt@M6pt{
n\times m	\ar[r]^{n\tensor \bar A} \ar[d]_{f\tensor m} & n \ar[d]^f\\
m	\ar[r]^{\bar A} & 1
}
\end{split}
\end{equation}
commutes. Now we have
\begin{align*}
  n\tensor\bar A & = \Brks{\bar A[\kappa_i\times\id_m]}_{i\in n}\\
  & = \Brks{\{[(\kappa_i\times\id_m)\kappa_j]\mid j\in A\}}_{i\in n}\\
  & = \Brks{\{[\kappa_{(i,j)}]\mid j\in A\}}_{i\in n}\\
  & = \{\Brks{[\kappa_{(i,j_i)}]}_{i\in n}\mid (j_i)\in A^n\}
\end{align*}
and hence
\begin{equation*}
  f(n\tensor\bar A) = \{f\Brks{[\kappa_{(i,j_i)}]}_{i\in n}\mid (j_i)\in A^n\}.
\end{equation*}
On the other hand, we have 
\begin{align*}
  \bar A(f\tensor m) & = \bar A\Brks{f[\id_n\times\kappa_j]}_{j\in m}\\
  & = \{[\kappa_j]\Brks{f[\id_n\times\kappa_j]}_{j\in m}\mid j\in A\}\\
  & \{f[\id_n\times\kappa_j\mid j\in A\}
\end{align*}
so that equivalence of the two sides follows from
\begin{equation}
  \label{eq:tensor-final}
  \{[\id_n\tensor\kappa_j]\mid j\in A\}\approx \{\Brks{\kappa_{(i,j_i)}}_{i\in n}
  \mid (j_i)\in A^n\}.
\end{equation}
To prove (\ref{eq:tensor-final}), we compare the projections along
$[\kappa_i]$, $i\in n$, on both sides, and thus reduce the goal to
the evident equality
\begin{align*}
  \{[\kappa_i][\id_n\times\kappa_j]\mid j\in A\}&=\{[\kappa_{(i,j)}\mid j\in A]\}\\&=
  \{[\kappa_{(i,j_i)}]\mid (j_i)\in A^n\}.
\end{align*}

We have thus shown that $T=T_0/\approx$ is a candidate for the tensor
product of $L$ and $\PSetN$. It remains to prove the universal
property. Thus, let $S$ be a further candidate, i.e.\ a Lawvere theory
with maps $G: L\to S$, $G_{\PSetN}:L_{\PSetN}\to S$ such that the
tensor law is satisfied for $G$ and $G_{\PSetN}$. We define a functor
$\bar G:T\to S$ as follows. In preparation, we note that every subset
$A\subseteq n$ of some set $n$ is a morphism $n\to 1$ in
$L_{\PSetN}$, whose image under $G_{\PSetN}$ we denote by
$(A\subseteq n):n\to 1$. In particular, for a morphism $A:n\to m$ in
$T_0$ we have $\hat A:=(A\subseteq L(n,m)):L(n,m)\to 1$ in
$S$. Moreover, we have for each set $n$ a morphism
\begin{equation*}
s_n=G(\Brks{f}_{f\in L(n)}):n\to L(n)
\end{equation*}
in $S$, where we denote by $L(n)$ the action of the monad induced by
$L$, i.e.\ simply $L(n)=L(n,1)$. We then define a functor $\bar
G_0:T_0\to S$ by putting, for $A:n\to 1$ in $T_0$,
\begin{equation*}
  \bar G_0(A)=\hat A s_n.
\end{equation*}
In general, we then put
\begin{equation*}
  \bar G_0(A)  = \Brks{G([\kappa_i]A)}_{i\in m}
\end{equation*}
for $A:n\to m$ in $T_0$ (noting that this agrees with the previous
definition in case $m=1$). 

To establish the requisite properties of $\bar G_0$, we need the
following lemma.
\begin{lem}\label{lem:image}
  Let $A\subseteq n$, $B\subseteq m$, and let $e:n\to m$ such that
  $e[A]=B$. Then
  \begin{equation*}
    (A\subseteq n)[e]=(B\subseteq m)
  \end{equation*}
  in $S$.
\end{lem}
\begin{IEEEproof}
  Immediate from the corresponding equality in $L_{\PSetN}$. 
\end{IEEEproof}

To begin, we now show that $\bar G_0$ preserves $[\argument]$, which
will then also imply that $\bar G_0$ preserves identities. Thus, let
$e:n\to m$ be a map; we have to show that $\bar G_0[e]=[e]$, which by
applying product projections on both sides and by definition of $\bar G_0$
immediately reduces to the case $n=1$, i.e.\ $e=\kappa_j$ for some
$j\in m$. Now we have
\begin{align*}
  \bar G_0[\kappa_j] & = \widehat{\{[\kappa_j]\}}s_m \\
  & = (1\subseteq 1)[\kappa_{[\kappa_j]}]s_m\\
  & = \kappa_j,
\end{align*}
where the second step is by Lemma~\ref{lem:image} (applied to
$\kappa_{[\kappa_j]}[1]=\{[\kappa_j]\}$).

The crucial point in the proof is now to establish that $\bar G_0$
preserves composition. Again, this reduces immediately to the case
where the codomain of the composite is $1$. Thus, let $A:n\to 1$, and
let $B:k\to n$ in $T_0$; put $B_j=[\kappa_j]B$ and $\nu_j=\lambda
g.\,[\kappa_j]g:L(k,n)\to L(k)$ for $j\in n$. By
Lemma~\ref{lem:image}, we then have $\hat B_j=\hat B[\nu_j]$. We start
to transform $\bar G_0A\bar G_0B$:
\begin{align*}
  \bar G_0A\bar G_0B&=\hat A s_n\Brks{\hat B_js_k}_{j\in n}\\
  & = \hat A s_n\Brks{\hat B[\nu_j]s_k}_{j\in n}\\
  & = \hat A s_n (n\tensor\hat B)\Brks{[\nu_j]s_k}_{j\in n}\\
  & = \hat A (L(n)\tensor \hat B)(s_n\tensor L(k,n))\Brks{[\nu_j]s_k}_{j\in n},
\end{align*}
using the tensor law in the last step. We proceed to analyse the right-hand
subterm of the last term separately: we claim that
\begin{equation}
  \label{eq:double-generic}
  s_n\tensor L(k,n))\Brks{[\nu_j]s_k}_{j\in n}=G\Brks{fg}_{f\in L(n),g\in L(k,n)}.  
\end{equation}
We then note moreover that the right-hand side of
(\ref{eq:double-generic}) equals $[c]s_k$, where $c:L(n)\times
L(k,n)\to L(k)$ is composition (this is proved by precomposing both
sides with the projections $[\kappa_{(f,g)}]$: we have
$[\kappa_{(f,g)}][c]s_k=[c\kappa_{(f,g)}]s_k=[\kappa_{fg}]s_k=G(fg)$.)
We then conclude the argument by
\begin{multline*}
  \hat A (L(n)\tensor \hat B)(s_n\tensor L(k,n))\Brks{[\nu_j]s_k}_{j\in n}\\ =
  \widehat{A\times B}[c]s_k  = \widehat{AB}s_k,
\end{multline*}
again using Lemma~\ref{lem:image} in the last step.

It remains to prove our claim (\ref{eq:double-generic}). We note that
$n\times L(k,n)$ is the $n$-fold product of $L(k,n)$ in $S$, with
projections $p_j=[\lambda g.\,(j,g)]$ for all $j\in n$, and at the
same time the $L(k,n)$-fold product of $n$, with projections
$q_g=[\lambda j.\,(j,g)]$ for all $g\in L(k,n)$, and similarly
$L(n)\times L(k,n)$ is the $L(k,n)$-fold product of $L(n)$, with projections
$\bar q_g=[\lambda f.\,(f,g)]$ for all $g\in L(k,n)$. We then prove (\ref{eq:double-generic}) by
precomposing both sides with $[\kappa_{(f,g)}]$. We have
\begin{align*}
&  [\kappa_{(f,g)}](s_n\tensor L(k,n))\Brks{[\nu_j]s_k}_{j\in n}\\
  & = [\kappa_f][\lambda f.\,(f,g)](s_n\tensor L(k,n))\Brks{[\nu_j]s_k}_{j\in n} \\
  & = [\kappa_f]\bar q_g(s_n\tensor L(k,n))\Brks{[\nu_j]s_k}_{j\in n} \\
  & = [\kappa_f]s_n q_g \Brks{[\nu_j]s_k}_{j\in n}\\
  & = G(f) q_g \Brks{[\nu_j]s_k}_{j\in n}.
\end{align*}
Thus we are done once we show that $q_g \Brks{[\nu_j]s_k}_{j\in
  n}=G(g)$. To this end, we precompose both sides with $[\kappa_j]$
and calculate
\begin{align*}
  & [\kappa_j]q_g\Brks{[\nu_j]s_k}_{j\in n}\\ 
& =  [\kappa_j][\lambda j.\,(j,g)]\Brks{[\nu_j]s_k}_{j\in n}\\
  & = [\kappa_g][\lambda g.\,(j,g)]\Brks{[\nu_j]s_k}_{j\in n}\\
  & = [\kappa_g]p_j\Brks{[\nu_j]s_k}_{j\in n}\\
  & = [\kappa_g][\nu_j]s_k \\
  & = [\kappa_{[\kappa_j]g}]s_k = G([\kappa_j]g)=[\kappa_j]Gg.
\end{align*}
This concludes the proof that $\bar G_0$ preserves composition. It is
then clear that $\bar G_0$ factors through $T$, as its kernel
satisfies all properties featuring in the inductive definition of
$\approx$ (the kernel is, of course, reflexive and transitive, and it
is closed under $(\pi)$ because tupling is unique in $S$). Uniqueness
of the arising factorizing morphism $\bar G:T\to S$ is clear, because
every morphism $n\to m$ in $T$ has the form $F_{\PSetN}(A) F(B)$, where
$B:n\to k$ in $L$ and $A:k\to n$ in $L_{\PSetN}$.

Summarizing the above, we have shown that
\begin{quote}
  \emph{$T_0/\approx$ is the tensor product of $L$ and $\PSetN$.}
\end{quote}
\bigskip
One consequence of this is the following property:
\begin{lem}
  $\approx$ is congruent w.r.t.\ union.
\end{lem}

\subsection{A Direct Construction of the Power Tensor}

\noindent We proceed to give details for Theorem~\ref{thm:p_char}.
The construction of tensor products with the full powerset theory
$L_\PSet$ is similar to the one for the nonempty powerset, but more
involved due to the fact that the full powerset theory has a constant,
$\emptyset$. The general construction of tensoring a theory $L$ with
uniform theories tells us that for such a case, we have to expect a
three-layered normal form that has operations of $L_\PSet$ on top,
under this a layer of operations of $L$, and at the bottom a layer
consisting not only of variables but possibly also of occurrences of
$\emptyset$.

To simplify matters, we have assumed that the given theory $L$ is
bounded, with the unique constant denoted $\bot$. This will in
particular allow us to replace occurrences of $\emptyset$ in the
bottom layer with $\bot$, thus effectively reverting to a two-layered
structure.

Under the assumption that $L$ is bounded, the tensor $L\tensor
L_\PSet$ is constructed as follows. As in the case of the nonempty
power tensor, we begin by constructing a preliminary category
$T_0$. Morphisms $n\to m$ in $T_0$ are (possibly empty) subsets of
$L(n,m)$. Composition is pointwise, as previously; also, the
definition of the embedding functors $L\to T_0$, $L_\PSet\to T_0$ from
the component theories remains unchanged, similarly for the indexing
functor $\Set\to T_0$. The crucial difference with the nonempty
powerset theory is that we have to adapt the definition of tupling to
work around the basic fact that Cartesian products of sets are empty
if one of the sets is empty, i.e.\ Cartesian products do not directly
provide a faithful representation of tuples of sets as sets of
tuples. Here, we exploit the fact that $\bot$ is available, and by the
tensor law is equivalent to the empty set. We thus put, for $A:n\to m$
in $T_0$,
\begin{equation*}
  \tilde A =
  \begin{cases}
    \{\bot\} & \text{if $A=\emptyset$}\\
      A & \text{otherwise}
  \end{cases}
\end{equation*}
and then define the tupling of a family of morphisms $A_i:n\to m$,
$i\in k$, by
\begin{equation*}
  \Brks{A_i}_{i\in k}=\{\Brks{f_i}_{i\in k}\mid \forall i\in k.\, 
  f_i\in\tilde A_i\}.
\end{equation*}
By the calculation carried out for the nonempty case, we then have
$\pi_j\Brks{A_i}=\tilde A_j$ in $T_0$, and subsequent quotienting will
ensure that $\tilde A_j$ becomes equal to $A_j$.  Commutation of the
tensor injections with the indexing functors is as in the nonempty
case. 

Next, we quotient $T_0$ by \emph{rectangular equivalence},
i.e.\ the relation $\poweq$ defined inductively as the smallest
reflexive and transitive relation closed under rule $(\pi)$ and
additionally satisfying the axiom
\begin{equation*}
  (\bot)\quad\{\bot_{n,m}\}\poweq\emptyset
\end{equation*}
as well as the symmetric $\emptyset\poweq\{\bot_{n,m}\}$ for all
$n,m$. We have

\begin{lem}
  The relation $\poweq$ satisfies the following properties.
  \begin{sparenumerate}
  \item $\poweq$ is symmetric.
  \item\label{item:tilde-poweq} $A\poweq\tilde A$ for all $A$.
  \item $\poweq$ is a congruence w.r.t.\ composition.
  \item $\poweq$ is a congruence w.r.t.\ tupling.
  \end{sparenumerate}
\end{lem}
\begin{IEEEproofNoQED}
  \begin{sparenumerate}
  \item As in the nonempty case.
  \item Trivial.
  \item Let 
    \begin{equation*}
      A\poweq^c B\iff \forall L,R.\,LAR\poweq LBR.
    \end{equation*}
    As in the nonempty case, $\poweq^c$ is easily seen to be
    reflexive, transitive, and closed under ($\pi$). It remains to see
    that $\poweq^c$ contains all instances of $(\bot)$, i.e.\
    (considering only one of the two symmetric cases of $(\bot)$) that
    for all $L,R$, $L\{\bot\}R\poweq L\emptyset R=\emptyset$. But this
    follows from boundedness of $L$: for $l\in L$, $r\in R$, we have
    $l\bot r=\bot$, and therefore
    $L\{\bot\}R=\{\bot\}\poweq\emptyset$. Therefore, $\poweq^c$
    contains $\poweq$, so that $\poweq$ is a congruence.
  \item Let $A_i,B_i:n\to m_i$ and $A_i\poweq B_i$ for $i\in k$. To
    prove $\Brks{A_i}\poweq\Brks{B_i}$, we have to show
    $[\kappa_{(i,j)}]\Brks{A_i}\poweq[\kappa_{(i,j)}]\Brks{B_i}$ for
    $i\in k$, $j\in m_i$. Now
    $\kappa_{(i,j)}=(\kappa_j\times\id_k)\iota_i$, where
    $\iota_i:m_i\to\sum m_i$ is the coproduct embedding. Thus, we have
    $[\kappa_{(i,j)}]\Brks{A_i}=[\kappa_j]\tilde
    A_i\poweq[\kappa_j]A_i\\poweq[\kappa_j]B_i\poweq[\kappa_j]\tilde
    B_i= [\kappa_{(i,j)}]\Brks{B_i}$, using the assumption $A_i\poweq
    B_i$, congruence w.r.t.\ composition, and
    claim~\ref{item:tilde-poweq} of this lemma. \myqed
  \end{sparenumerate}
\end{IEEEproofNoQED}

So far, we have established that $T:=T_0/\poweq$ is a large Lawvere
theory that has theory morphisms $L\to T$ and $L_\PSet\to T$. To prove
that $T$ is a candidate for the tensor $L\tensor L_\PSet$, we need to
show that the tensor law holds. The argument is mostly as in the
nonempty case: we need only check those cases where the empty set can
occur within a tupling operation; the only case in point is where
$A=\emptyset$, in the notation of (\ref{eq:tensor}). This case,
however, is taken care of by the fact that $\emptyset\poweq\{\bot\}$
and by boundedness of $L$, which ensures that for $\{\bot\}$ in place
of $\bar A$, both paths in (\ref{eq:tensor}) equal $\{\bot\}$.

It remains to prove the universal property. Given a further candidate
$S$, i.e.\ a large Lawvere theory with morphisms $G:L\to S$ and
$G_\PSet:L_\PSet\to S$ satisfying the tensor law, we define $\bar
G_0:T_0\to S$ as before; the proof that $\bar G_0$ respects
composition and $[\argument]$ is unchanged from the nonempty
case. Again, it is clear that $\bar G_0$ factors through $T$ because
its kernel satisfies the inductive definition of $\poweq$, including
all instances of $(\bot)$ as these are implied by validity of the
tensor law in $S$. Uniqueness of the factorization is, again, clear.

\subsection{Proof of Lemma~\ref{lem:p1_approx}}
\noindent We need a preliminary lemma:
\begin{lem}\label{lem:downup-cong}
  Let $A:n\to m$ in $T_0$. Then for all $a:n\to m$ and all $b:m\to k$,
  \begin{equation}
    a\in\downup{A}\implies ba\in\downup{bA}.
  \end{equation}
\end{lem}
\begin{IEEEproof}
  It suffices to show that the set $\bar A=\{a\in\downup{A}\mid
  ba\in\downup{bA}\}$ contains $A$, is downward closed, and is
  closed under ($\Delta$). The first and second properties are clear;
  we check the third property. Thus, let $h:n\to k$ and $g:k\to m$
  such that $g\Delta_jh\in\bar A$ for all $j\in k$. Then
  $bg\Delta_jh\in\downup{bA}$ for all $j$, and hence
  $bgh\in\downup{bA}$, so that $gh\in\bar A$ as required.
\end{IEEEproof}

\begin{IEEEproofNoQED}[Proof of Lemma~\ref{lem:p1_approx}]
  In preparation, note that
  \begin{equation}\label{eq:delta-alt}
    \Delta_i=\Brks{\delta_{ij}}_{j\in n}[\kappa_i].
  \end{equation}
  \emph{Only if:} It suffices to show that the equivalence $\simeq$
  defined by $A\simeq B$ iff $\downup{A}=\downup{B}$ is closed under
  $(\pi)$ and contains all instances of $(\bot)$. The latter holds by
  the definition of $\downup{\emptyset}$. To check the former, we
  first show that $\simeq$ is left congruent w.r.t.\
  composition. Thus, let $A,B:n\to m$ and let $C:m\to k$ such that
  $\downup{A}=\downup{B}$. We have to show
  $\downup{CA}=\downup{CB}$. Since $\downup{CB}$ is downward
  closed and closed under ($\Delta$), it suffices to prove
  $CA\subseteq\downup{CB}$. Thus let $c\in C,a\in A$. Then
  $a\in\downup{B}$ by assumption, and therefore
  $ca\in\downup{cB}\subseteq\downup{CB}$ by
  Lemma~\ref{lem:downup-cong}). It follows that
  $\downup{CA}\subseteq\downup{CB}$. The converse implication is
  shown symmetrically.

  It remains to show that $\simeq$ is closed under ($\pi$). Thus, let
  $A,B:n\to m$ such that $\downup{[\kappa_i]A}=\downup{[\kappa_i]B}$
  for all $i\in m$. We have to show $\downup{A}=\downup{B}$. By
  \eqref{eq:delta-alt} and Lemma~\ref{lem:downup-cong}, we have
  $\Delta_ia\in\downup{(\Delta_iB)}$ for all $i\in m,a\in A$. By
  downward closedness, $\downup{(\Delta_iB)}\subseteq\downup{B}$, so
  that we obtain $a\in\downup{B}$ by rule ($\Delta$). It follows that
  $\downup{A}\subseteq\downup{B}$; the reverse inclusion is shown
  symmetrically.

  \emph{If:} It suffices to show that for $A:n\to m$ in $T_0$,
  $A\poweq\downup{A}$. Since $\poweq$ is congruent w.r.t.\ set
  union, it suffices to show that $A\poweq A\cup\{f\}$ for all
  $f\in\downup{A}$, which will follow if we show that the set 
  \begin{equation*}
    \bar A:=\{f:n\to m\mid A\poweq A\cup\{f\}\}
  \end{equation*}
  (which clearly contains $A$) is downward closed and closed under $\Delta$.
  \begin{sparitemize}
  \item \emph{$\bar A$ is downward closed:} define a preorder $\unlhd$
    by $f\unlhd g:\iff\{f,g\}\poweq g$; then $\unlhd$ is easily
    seen to be closed under $\pi_{\sqsubseteq}$, and hence contains
    $\sqsubseteq$. Now let $g\in\bar A$, $f\sqsubseteq g$. Then $f\unlhd g$ and therefore
    \begin{equation*}
      A\cup\{f\}\poweq A\cup\{g\}\cup\{f\}\poweq A\cup\{g\}\poweq A,
    \end{equation*}
    using congruence of $\poweq$ w.r.t.\ union.
  \item \emph{$\bar A$ is closed under ($\Delta$):} Let $f:n\to m$,
    $g:m\to k$, and let $f\Delta_i g\in\bar A$ for all $i\in m$. Then
    \begin{equation*}
      A\cup\{fg\}\poweq A\cup\{f\Delta_i g\mid i\in m\}
      \poweq A,
    \end{equation*}
    using congruence w.r.t.\ union and Lemma~\ref{lem:delta}. \myqed
  \end{sparitemize}
\end{IEEEproofNoQED}

\subsection{Proof of Lemma~\ref{lem:delta}}
\noindent By rule ($\pi$), it suffices to prove
$[\kappa_j]g\poweq[\kappa_j]\{\Delta_ig\mid i\in m\}$ for all $j\in
m$. But the right hand side of this equivalence equals
$\{[\kappa_j]g\}\cup\{\bot\}$, which is equivalent to $[\kappa_j]g$ by
($\bot$) and congruence w.r.t.\ union.\myqed

\subsection{Proof of the Order-Theoretic Conservativity
  Theorem~\ref{thm:ord-cons}}
\begin{sparenumerate}
  \item \emph{a) $\impl$ b):} Immediate from Lemma~\ref{lem:delta}.

    \emph{b) $\impl$ c):} The inclusion $\downup{f}\subseteq
    f\downset$ holds because under b), $f\downset$ is closed under
    ($\Delta$). The reverse inclusion holds because $\downup{f}$ is
    downclosed.

    \emph{c) $\impl$ a):} Immediate by Lemma~\ref{lem:p1_approx}.
  \item Immediate from property c). \myqed
  \end{sparenumerate}

\subsection{Details for Remark~\ref{rem:eq-cons}}
\noindent Let $L$ be simply ordered. We prove that if $\sigma_1:L\to
L\tensor L_\PSet$ is faithful, then it reflects the
ordering. According to Theorem~\ref{thm:ord-cons}, we have to show
that for $f:n\to m$, $g:m\to k$ in $L$, $fg=\bigsqcup_{i\in
  m}f\Delta_ig$. Thus let $h$ be an upper bound of $\{f\Delta_ig\mid
i\in m\}$. Since $L$ is simply ordered, there exist $f'$, $g'$ such
that $f' g'\sqsubseteq h$, $f' g'$ is a minimal upper bound of
$\{f\Delta_i g\mid i\in m\}$, and for every $j$ there is $i$ such that
$f'\Delta_j g'\sqsubseteq f\Delta_i g\sqsubseteq f g$. By $(\Delta)$,
$f' g'\in \downup{f g}$ and $f g\in \downup{f' g'}$. Hence $\downup{f
  g}=\downup{f' g'}$. Faithfulness of $\sigma_1$ then implies that $f
g=f' g'\sqsubseteq h$.

\subsection{Details for Example~\ref{expl:cons}}

\noindent \emph{Absolutely free theories}: For the sake of
readability, we restrict to the case where $L$ is generated by a
finitary signature $\Sigma$ (and, of course, no equations). Then
$L\tensor L_\bot$ is generated by $\Sigma$ and an additional constant
$\bot$, and equations
\begin{equation*}
  f(\bot,\dots,\bot)=\bot
\end{equation*}
for every basic function symbol $f$ in $\Sigma$. We can direct these
equations from left to right and obtain a single-step rewrite relation
$\to$ which is clearly strongly normalizing and trivially locally
confluent, hence confluent by Newman's lemma. We define a syntactic
approximation $\sqsubseteq_0$ on $\Sigma\cup\{\bot\}$-terms by
$t\sqsubseteq_0 s$ iff there exist variables $x_1,\dots,x_n$ such that
$t=s[\bot/x_1,\dots,\bot/x_n]$; i.e.\ $t\sqsubseteq_0 s$ iff the term
$t$ is obtained from $s$ by deleting some subterms. Then
\newcommand{\NF}{\mathit{NF}}
\begin{claim}\label{claim:approx}
  $t\sqsubseteq s$ iff $\NF(t)\sqsubseteq_0 s$
\end{claim} 
\noindent where $\NF(t)$ denotes the normal form w.r.t.\ $\to$. `If'
is trivial since $t\sqsubseteq_0s$ clearly implies $t\sqsubseteq s$.

To prove `only if', first note that $t\sqsubseteq s$ iff $s$ can be
reached from $t$ by a chain of terms $t=t_0,\dots,t_n=s$ such that for
each $i=0,\dots,n-1$, either $t_i\sqsubseteq_0 t_{i+1}$ or
$t_i=t_{i+1}$ modulo the equations. We now show that whenever
$t\sqsubseteq_0 s\leftrightarrow s'$, then there exists $t'$ such that
$t\to^* t'\sqsubseteq_0 s'$, where $\leftrightarrow$ denotes the
symmetric closure of $\to$, and $\to^*$ the transitive reflexive
closure. Since $s'\to s$ implies $s\sqsubseteq_0 s'$ and
$\sqsubseteq_0$ is clearly a partial order, it suffices to consider
the case $s\to s'$, that is, $s'$ is obtained from $s$ by deleting an
occurrence $r$ of a subterm of the form $f(\bot,\dots,\bot)$. If $r$
is still present in $t$, we can delete it in $t$, obtaining $t'$ such
that $t\to t'$; then clearly $t'\sqsubseteq_0 s'$. If $r$ is deleted
in $t$, then already $t\sqsubseteq_0 s'$.

Now it follows that $t\sqsubseteq s$ iff there exists $t'$ such that
$t\to^*t'\sqsubseteq s$. We are done by noting that
$\NF(t)=\NF(t')\sqsubseteq_0 t'$. This proves
Claim~\ref{claim:approx}.

We now set out to prove the equality $\bigsqcup_{i\in m}
f\Delta_ig=fg$ for $g:n\to m$, $f:m\to k$. Translating this into the
language of $\Sigma\cup\{\bot\}$-terms, we have to show for a term $t$
in variables $x_,\dots,x_m$ and terms $s_i$ that $\bigsqcup
t\sigma_i=t\sigma$, where the substitutions $\sigma$, $\sigma_i$ are
defined by $\sigma(x_i)=s_i$ and by $\sigma_i(x_i)=s_i$,
$\sigma_i(x_j)=\bot$ for $i\neq j$, respectively. Thus,
$t\sigma_i\sqsubseteq_0 t\sigma$ for all $i$, i.e.\ $t\sigma$ is an
upper bound. Let $h$ be a further upper bound of the $t\sigma_i$
w.r.t.\ $\sqsubseteq$. By Claim~\ref{claim:approx},
$\NF(t\sigma_i)\sqsubseteq_0 h$ for all $i$. We are done once we prove
that $\NF(t\sigma)$ is the least upper bound of the $\NF(t\sigma_i)$
w.r.t.\ $\sqsubseteq_0$. This just means that a subterm at position
$r$ in $t\sigma$ is deleted in $\NF(t\sigma)$ iff it is deleted in all
of the $t\sigma_i$. Here, `only if' is clear (since the $\sigma_i$
only introduce more occurrences of $\bot$). To prove `if', we proceed
as follows. If $r$ lies within a substituted occurrence of $x_i$,
i.e.\ within an occurrence of $s_i$, then the subterm at position $r$
is the same in $t\sigma$ as in $t\sigma_i$, and we are
done. Otherwise, the position $r$ lies within $t$. Our goal then
reduces to showing that for every term $q$ in the variables $x_i$,
$q\sigma_i=\bot$ (modulo the equations) for all $i$ implies
$q\sigma=\bot$. We prove this by induction on the term structure. The
base case (where $q$ is one of the variables $x_i$) is clear. If $q$
is of the form $f(p_1,\dots,p_w)$, then $q\sigma_i=\bot$ implies (by
confluence of $\to$) that $p_j\sigma_i=\bot$ for all $i,j$, so that
$p_j\sigma=\bot$ for all $j$ by induction, and hence $q\sigma=\bot$.

\emph{Finitely additive finitary Lawvere theories:} Let $L$ be a
finitely additive finitary Lawvere theory; we prove that $L$ admits
unbounded nondeterminism. Recall that by Lemma~\ref{lem:add-order},
the approximation ordering on $L$ coincides with the ordering induced
by the additive structure. By Theorem~\ref{thm:ord-cons}, all that
remains to be shown is that for all $f:n\to m$, $g:m\to k$ in $L$,
$fg=\sum_{j\in m} f\Delta_j g$, where we can restrict to finite
$n,m,k$ because $L$ is finitary. Since finite sums commute with
composition, this reduces to showing that $\sum_{j\in
  m}\Delta_j=\id_m$, which is straightforward by comparing
projections.
\end{document}
